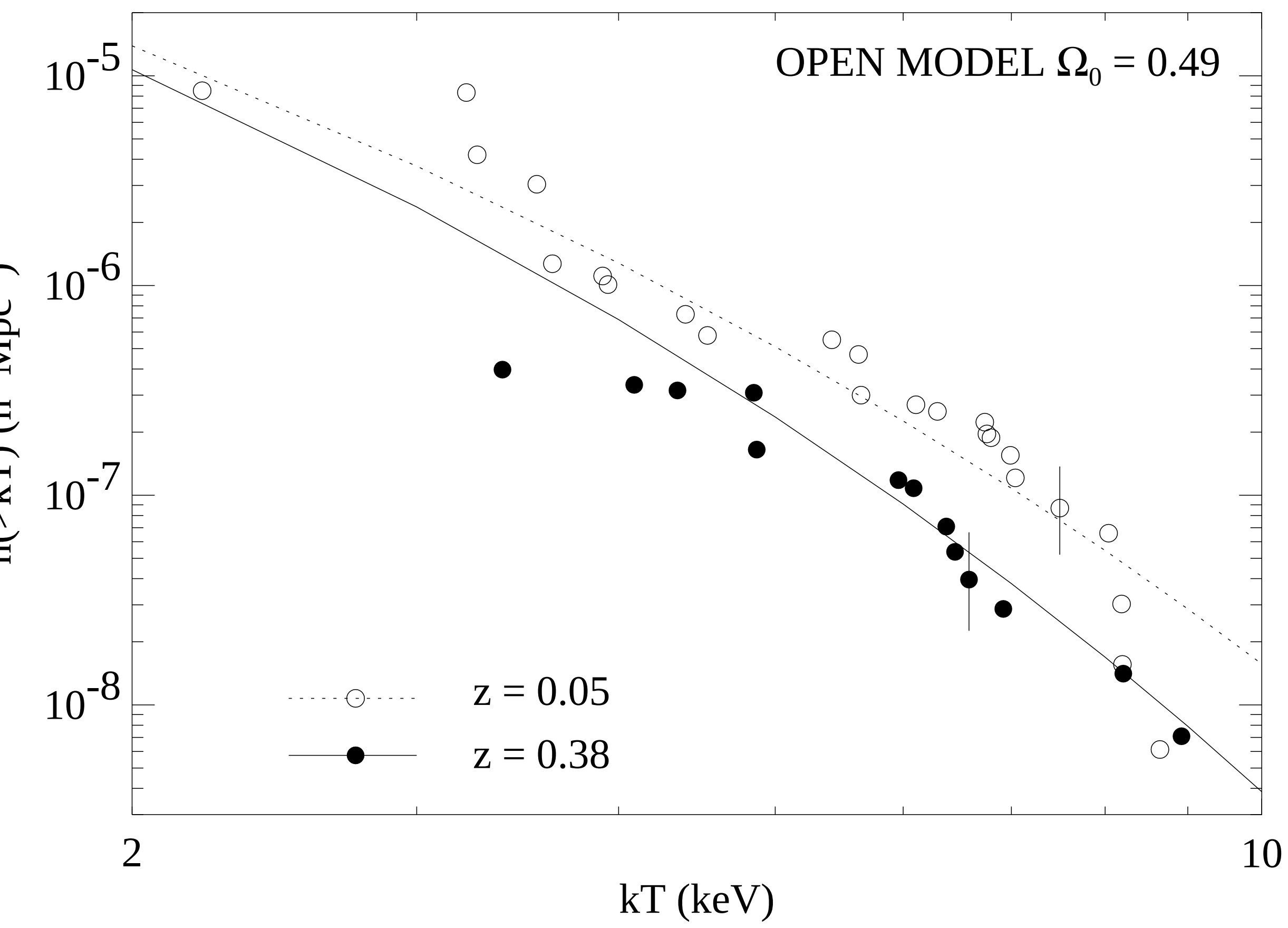

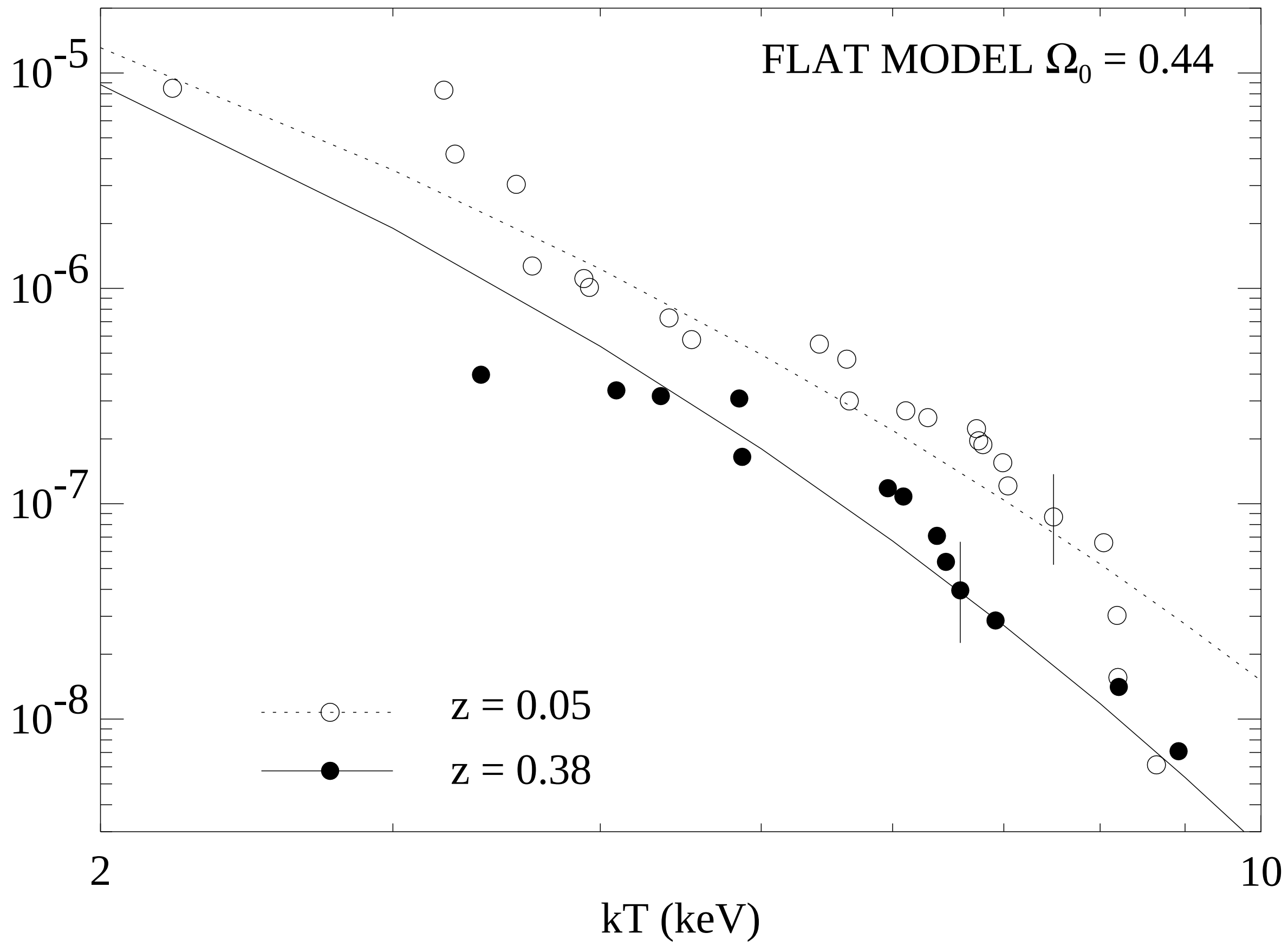

MEASURING COSMOLOGICAL PARAMETERS FROM THE EVOLUTION OF CLUSTER X-RAY TEMPERATURES

J. Patrick Henry
Institute for Astronomy, 2680 Woodlawn Dr., Honolulu, HI 96822

ABSTRACT

We have determined the cluster X-ray temperature function from two flux- and redshift-limited samples of clusters. The first sample is comprised of 25 clusters with average redshift 0.05. The local temperature function derived from it supercedes the one we previously published (Henry & Arnaud, 1971). Fourteen clusters with average redshift 0.38 comprises the second sample. We perform maximum likelihood fits of cluster evolution models to these data in order to constrain cosmological parameters. For an open model with zero cosmological constant we find that the density parameter is $\Omega_0 = 0.49 \pm 0.12$, the rms mass density fluctuation averaged over 8 $h^{-1}$ Mpc spheres is $\sigma_8 = 0.72 \pm 0.10$, and the effective index of the mass density fluctuation spectrum on cluster scales is $n = -(1.72 \pm 0.34)$ where all errors are symmetrized at 68% confidence. The corresponding results for the case where a cosmological constant produces a flat universe are: $0.44 \pm 0.12$, $0.77 \pm 0.15$, and $-(1.68 \pm 0.38)$. These results agree with those determined from a variety of different independent methods, including supernovae, galaxy-galaxy correlations, fluctuations in the microwave background, gravitational lens statistics, and cluster peculiar velocities.

## 1. INTRODUCTION

The properties of clusters of galaxies are rather sensitive to several parameters of cosmological interest. One of the first to be constrained was $\sigma_8$, the amplitude of mass density fluctuations on a scale of 8 $h^{-1}$ Mpc. Here, h is the current value of the Hubble parameter $H_0$ in units of 100 km $s^{-1}$ $Mpc^{-1}$. Using the number density of nearby clusters as a function of the velocity dispersion of their member galaxies, Evrard (1989) found a value between 0.67 and 1, Cole & Kaiser (1989) found $0.5 \pm 0.1$ and Peebles, Daly & Juszkiewicz (1989) found ~ 0.67. Henry and Arnaud (1991, HA91) were the first to use cluster X-ray temperatures to measure $\sigma_8$, finding $0.59 \pm 0.02$ from a local sample.

For reasons of simplicity, not to mention fashion, these early analyses were restricted to $\Omega_0 = 1$, where $\Omega_0$ is the ratio of the mass density of the universe at the present epoch to the critical density required to halt the expansion. More recent work has relaxed that assumption, showing that a local sample can only constrain a combination of $\Omega_0$ and $\sigma_8$. A large number of papers have performed these kinds of analyses using cluster abundances as a function of velocity dispersion, temperature, or mass (deduced from either the previous two observables or from gravitational lensing), or the logN - logS number counts (White, Efstathiou, & Frenk 1993; Eke, Cole, & Frenk, 1996; Viana & Liddle 1996; Kitayama & Suto, 1997; Mathiesen & Evrard, 1998 among others). The result of all this work is that $\sigma_8$ is in the range of 0.5 to 0.6, with errors ranging from 5% to 20% at 68% confidence, if $\Omega_0 = 1$. As $\Omega_0$ becomes less than one $\sigma_8$ increases, in order that the number of peaks above a given threshold remains at the observed value. It is worth noting that the value $\sigma_8$ is reasonably robust, having remained close to the above range since the very first determinations.

Additional information is required to break the degeneracy between $\sigma_8$ and $\Omega_0$. One source of that new information is the cosmic microwave background fluctuations. These fluctuations are currently measured on a vastly different size scale from clusters and some method has to be used to bridge the gap with its associated error. Another source of additional information is the evolution of cluster number density, which has the advantage that the new data pertain to the same size scale as the old. It was realized early on that cluster evolution was extremely sensitive to $\Omega_0$ (White and Rees, 1978; Perrenod, 1980; Evrard, 1989; Peebles et al. 1989; Frenk et al. 1990; Lilje, 1992; Oukbir & Blanchard 1992). The few high redshift clusters known, as well as instrumental difficulties, limited the early application of this technique using X-ray observations (Perrenod & Henry, 1981; Henry et al, 1982). Cluster evolution is not as sensitive to the cosmological constant, parameterized by $\lambda_0$, the present ratio of the cosmological constant $\Lambda$ to 3 times the square of $H_0$ (Eke et al, 1996; Viana & Liddle, 1996).

Recently, a number of papers have appeared that constrain $\Omega_0$ through a comparison of the properties of high and low redshift clusters (Carlberg et al, 1997; Bahcall, Fan & Cen 1997; Fan, Bahcall & Cen, 1997; Henry, 1997, H97; Sadat, Blanchard & Oukbir, 1998; Eke et al, 1998; Viana & Liddle, 1999; Borgani et al, 1999; Reichart et al, 1999; Donahue & Voit, 1999; Kay & Bower 1999). There is a wider divergence of results compared to the situation with $\sigma_8$. With the exception of Sadat et al (1998) and Reichart et al (1999), all this work is consistent with a low density universe ($0.2 < \Omega_0 < 0.5$) with $\Omega_0 = 1$ excluded at varying levels of significance ranging from less than one sigma to greater than $10^{-6}$! Some of this discrepancy is surely due to the different redshift ranges probed; higher redshift objects have more power to discriminate. Almost just as surely, however, are uncorrected systematic biases, either observational or theoretical.

The shape of the mass fluctuation spectrum may be measured as well. Early work showed that the standard cold dark matter model was inconsistent with the cluster distribution, either as a function of temperature (HA91; Oukbir & Blanchard, 1997), X-ray luminosity and its evolution (Henry et al, 1992) or of mass derived from temperatures, and/or velocity dispersions (Bahcall & Cen 1992). Rather a steeper spectrum (more negative spectral index) was preferred. Constraining the shape of the fluctuation spectrum has not received as much attention as the other cosmological parameters, with most workers assuming a cold dark matter spectrum with shape parameter measured from galaxy surveys. However, cluster data provide independent constraints that agree with those from galaxies.

We have previously given a brief report (H97) of the constraints on the above cosmological parameters provided by the cluster X-ray temperature function and its evolution. Using the same data, we have also investigated the effects of systematic uncertainities on these constraints (Eke et al, 1998). In this paper we improve on H97 as well as provide all the details of its analysis. Among the improvements are new X-ray temperatures from ASCA for the low redshift sample, and an enlarged high redshift sample for which the temperatures of all clusters have been derived uniformly from the same analysis using the latest ASCA calibrations. We also compare our results with those from independent constraints.

## 2. OBSERVATIONAL RESULTS

Our primary observation goal is to construct the cluster temperature functions at low and high redshifts. Since rich clusters have a wide range of temperatures, from ~2 keV to ~15 keV, we can only measure any evolution by comparing the temperature distributions. There are no standard temperature clusters.

The observed temperature function is a quite steep function of temperature (approximately $[kT]^{-5}$) and it is difficult to show it graphically with small sized samples. After experimenting with several methods, we have decided to present our data in the form of the integral temperature function, $n(>kT)$. This form has the advantage that it is nonparametric and unbinned. Its disadvantage is that the error on any point is difficult to establish since all points but the hottest are correlated. However, because of the very steep dependence on temperature, even a factor of two difference in temperature is enough to insure virtually uncorrelated data Thus we show representative error bars calculated from the Poisson noise on the number of clusters used to determine a given point on the distribution. The previous remarks refer to the graphical representation of our results. In this paper we will always use a maximum likelihood fitting technique to the individual unbinned measurements that properly calculates the errors on the fit parameters.

The integral temperature function is estimated from the data using

$$n(>kT) = \sum_i \frac{1}{V_{sea,i}} \tag{1}$$

where the sum is over all clusters with $kT_i > kT$ and $V_{sea,i}$ is the search or maximum volume in which the $i^{th}$ cluster could have been detected given its selection criteria.

## 2.1 LOW REDSHIFT SAMPLE

The low redshift sample remains the 25 clusters from HA91. We give the properties of these objects in Table 1. This sample came originally from the early all-sky surveys with non-imaging detectors. Confusion has been reduced by using pointed observations, still with non-imaging detectors, but with greatly reduced fields of view compared to the survey instruments. All but the Coma and Virgo clusters are unresolved with these instruments, so accurate values of the total flux are available.

The selection criteria are that the total flux in the 2 - 10 keV band is greater than 3 x $10^{-11}$ erg $cm^{-2}$ $s^{-1}$, the redshift is less than 0.17, and the absolute galactic latitude is greater than $20^o$ (plus a small zone of exclusion around the Magellanic Clouds) yielding a solid angle surveyed of 8.23 steradians. The redshift restriction was designed to insure a local sample. Its precise value was set to exclude clusters with temperatures greater than 10 keV through the luminosity - temperature relation (see Section 3). The maximum redshift at which a cluster could have been detected is given to an adequate approximation by the Euclidean limit

$$z_{\max,i} = \left(\frac{F_i}{F_{\lim}}\right)^{1/2} z_i \tag{2}$$

where $F_{lim}$ is 3 x $10^{-11}$ erg $cm^{-2}$ $s^{-1}$. Use of the Euclidean expression for the search volume calcluated from $z_{max}$ introduces an error of less than 30%. Still we use the exact formula for an open model given in the appendix for the low redshift sample. This calculation is an improvement on our previous work. The difference between the search volumes for open and flat models is about 5% for this sample, which we neglect. We give the search volumes in Table 1 and the resulting integral temperature functions in the table and Figures 1 and 2. The median search volume for the low redshift sample is 2.9 x $10^7$ $h^{-3}$ $Mpc^3$.

We need to mention the case of A2163. This cluster satisfies all the selection criteria except its redshift is 0.201, thus it is excluded from our low redshift sample. Certainly, it would be difficult to argue that a cluster at such a redshift is local. At a temperature of $14.6^{+0.9}_{-0.8}$ keV it would have been the hottest in our sample. Its $z_{max}$ is 0.213 implying a search volume of 4.96 x $10^8$ $h^{-3}$ $Mpc^3$ (for $\Omega_0$ = 0.5), so if it were included in the sample, it would have added 2.02 x $10^{-9}$ $h^3$ $Mpc^{-3}$ to all points in the integral temperature function. This contribution is small, except for the hottest cluster in Table 1.

An important improvement over H97 is the recent availability of temperatures from ASCA measurements of nearly two-thirds of the sample (Markevitch et al 1998). We want to compare temperature functions at low and high redshifts, so we use the ASCA single temperature fits uncorrected for cooling flows from Markevitch et al (1998) because such corrections are not yet possible for the high redshift clusters. Until now the temperatures for the low and high redshift samples came from different instruments. These new measurements help to eliminate a possible source of systematic error. Fortunately for previous work, the ASCA and non-ASCA temperature scales are the same to a few percent, as is expected. The error-weighted mean ratio of ASCA to mean non-ASCA temperatures in Table 1 for clusters that can be compared is $1.031 \pm 0.022$ (68% confidence error on the mean).

Edge et al (1990), HA91, and Markevitch (1998) have previously reported the low redshift temperature function. Their differential temperature functions were fit to a power law, $n(kT) = A[kT]^{-\alpha}$. The best fitting A (in units of $h^3$ Mpc$^{-3}$ keV$^{-1}$) and $\alpha$ were: $8.8^{+7.2}_{-4.0} x10^{-4}$ and 4.93±0.37, $1.8^{+0.8}_{-0.5} x10^{-3}$ and 4.7±0.5, and $3.5^{+6.0}_{-2.2} x10^{-4}$ and 4.2±0.7 respectively. The power law indices agree within their errors. The normalizations are somewhat difficult to compare because of the steep temperature dependence. At a temperature of 6 keV, the differential temperature functions are (in units of $10^{-7}$ $h^3$ Mpc$^{-3}$ keV$^{-1}$) 1.28±0.08, $3.96^{+3.05}_{-1.62}$, and 1.90±0.54 respectively after allowing for the correlated errors in A and $\alpha$ (a larger $\alpha$ requires a larger A to maintain the total number of clusters).

This comparison shows that the HA91 normalization is systematically high, although only marginally so with respect to the Markevitch (1998) determination. Eke et al (1996) pointed out to us that this higher normalization was in error. Upon investigation, we found that the HA91 differential temperature function contained two nearly compensating errors. The first was a typographical error in a numerical constant in our code. The second was subtler (almost anything is subtler than a typo). A determination of the differential temperature function requires binning the data and performing an average over the objects in the bin. This procedure will introduce some arbitrariness that the use of the integral function avoids. There are two obvious averages that come to mind. The one used in HA91 was $1/\langle V_{sea,i}\, \Delta T_i \rangle$ where $\langle \ldots \rangle$ denotes average. The estimator commonly used is $\langle 1/(V_{sea,i}\, \Delta T_i) \rangle$. The former average is always less than the latter, unless all objects in the average are identical. Further, the contribution of a given cluster to the temperature function depends on that of other clusters with similar temperatures using the first average. In the second average, the contribution of any clusters to the temperature function depends only on its unique properties, which is clearly to be preferred.

We have, therefore, made a new fit of the low redshift temperature function to a power law using the maximum likelihood procedure described in section 4. The best fitting parameters are A = $1.0^{+2.5}_{-0.7} x10^{-3}$ $h^3$ Mpc$^{-3}$ keV$^{-1}$ and $\alpha$ = 4.9±0.7. This function is practically indistinguishable from that of Edge et al (1990) and Markevitch (1998).

## 2.2 HIGH REDSHIFT SAMPLE

The high redshift sample comes from the *Einstein* Extended Medium Sensitivity Survey (EMSS, Gioia et al 1990; Henry et al 1992, H92 hereafter; Gioia and Luppino 1994). The EMSS was conducted with an imaging detector; hence its fundamental variable is surface brightness. Sources in the EMSS were found serendipitously in fields targeted at unrelated objects. Each source was required to produce a sufficient number of counts in a 2.4' x 2.4' detect cell to be significantly above the expected background in that cell for the exposure of the field. The selection criteria to be included in the sample used here are that the detect cell flux in the 0.3 - 3.5 keV band be greater than 2.5 x $10^{-13}$ erg cm$^{-2}$ s$^{-1}$, $\delta(1950) \geq -40^{\circ}$, and the redshift satisfy $0.3 \leq z < 0.6$. There are sixteen clusters in Gioia and Luppino (1994) satisfying our selection criteria. However two of them, MS 1333.3 and MS 1610.4, are X-ray point sources (Molikawa et al, 1998; our unpublished analysis). We removed these two objects from our sample leaving a total of fourteen. All objects discussed here are extended X-ray sources in deep ROSAT HRI observations, that is they are all *bona fide* clusters. Finally we have revised the redshift of MS1241.5 based on new spectral observations taken at the Hawaii 2.2m telescope (Gioia, 1998 private communication). After these revision the high redshift sample has an average redshift of 0.38. We also give our measurements for MS2053.7 whose flux is just below the cutoff and thus not in the sample used to constrain cosmological parameters.

We have determined temperatures, fluxes, and luminosities for the high redshift clusters from ASCA observations. These data are given in Table 2. All clusters were analyzed identically using the latest calibrations. Specifically, we have used version 2 of the telescope effective area and point spread functions. The temperatures were hardly affected compared to H97 and Eke et al (1998). However, these files eliminate the ad hoc 18% correction to the Gas Imaging Spectrometer (GIS) flux employed there, although comparison of the two results shows that it was an excellent fudge factor. We have also taken care to use the 128 channel GIS detector 3 stuck bits Redistribution Matrix File (RMF) for MS 0451.6, the only cluster in our sample so affected. Spectra were extracted from 6.125' and 2.5' radii regions for the GIS and Solid-state Imaging Spectrometer (SIS) respectively. Background was acquired from adjacent regions of the same exposure. We grouped the spectra until each bin contained at least 20 counts, and then fit the two GIS and two SIS spectra simultaneously using the program XSPEC 10. The adjustable parameters were the GIS and SIS normalizations, temperature, abundance, and Hydrogen column density for the SIS (because of low energy calibration uncertainties). Fixed parameters were the redshift and GIS Hydrogen column density. The total flux and luminosity were determined from the GIS normalizations, since most of the observations were conducted in 2CCD mode and thus some of the photons were lost to gaps between the chips and off the edge of the detectors. We assumed that these objects are point sources in order to determine a total flux, a good approximation for distant clusters observed with the GIS on ASCA. The results are in Table 2. Our measurements of the temperature are in excellent agreement with previous values reported by others. The fluxes determined by ASCA in the 0.3 – 3.5 keV band are ~15% higher here than in H97 and Eke et al (1998) because they mistakenly did not correct for interstellar absorption. The 2 – 10 keV luminosities are hardly effected by this omission.

The standard EMSS procedure for determining the fraction, f, of a cluster's total flux in the EMSS detect cell is to integrate a beta model cluster surface brightness distribution, with $\beta$ = 2/3, over the detect cell and divide that result by the flux from the same beta model integrated to infinity, see equation (1) of H92:

$$f\left(\frac{\theta_D}{\theta_0}\right) = \frac{2}{\pi}\sin^{-1}\left(\frac{\theta_D^2/\theta_0^2}{\theta_D^2/\theta_0^2+1}\right) \tag{3}$$

where $\theta_D$ is the angular half-size of the detect cell, 1.2', and $\theta_0(z)$ is the angular size of the cluster core radius, $a_0/D_A(\Omega_0,z)$ with $D_A$ the angular diameter distance, see the Appendix, and $a_0$ the linear core radius. We use $a_0 = 0.125h^{-1}$ Mpc, which is appropriate for low (Mohr, Mathiesen & Evrard, 1999) and high (Vikhlinin et al, 1998) redshifts. The error weighted mean ratio of the ASCA to EMSS flux for all sources in Table 2 is 1.17 ± 0.05 (68% confidence error on the mean). The median value is 1.19. These values apply to either total or detect cell fluxes, since the same f from equation (3) converts either the EMSS detect cell to total or the ASCA total to detect cell. The factor ~2 discrepancy for MS0015.9, the largest deviation, was first pointed out by Nichol et al (1997) and may result from the object lying at the extreme edge of the field of view in the EMSS. Excluding this source gives a weighted average flux ratio of 1.12 ± 0.05 (68% confidence error on the mean) and a median of 1.15. We show a comparison of the fluxes of these clusters from the original EMSS and our new ASCA measurements in Figure 3.

The cause of this small discrepancy is not yet clear. We can think of a number of causes. First, it may simply be due to the absolute calibration uncertainties of the Einstein and ASCA observatories which are of the same order. Second, it could result from unrelated point sources in the GIS extraction beam that has a solid angle 20 times larger than the EMSS detect cell. To assess this possibility, we have examined the HRI images of all sources in Table 2. There are five objects (MS2137.3, MS1241.5, MS1358.4, MS0353.6, and MS1224.7) that do not have obvious point sources in the GIS extraction beam. All others have one or more such sources. The error weighted mean ratio of the ASCA to EMSS flux for these five is $0.99 \pm 0.08$ (68% confidence error on the mean). The median value is 1.09. Therefore, we regard not being able to exclude unrelated sources in the much larger GIS beam as the likely cause of the discrepancy.

However, Jones et al (1998) and Ebeling et al (1999) note a somewhat larger discrepancy in the same sense between the EMSS and ROSAT PSPC and HRI determined fluxes respectively and so we investigated further. One effect not included in equation (3) is the IPC point spread function. Including the effects of the point spread function on f, $f_{PSF}$, is substantially more complicated. In the end, $f/f_{PSF}$ is nearly a constant at all redshifts. For $z > 0.3$, it is 1.373 to better than 0.3%; for all redshifts for which the EMSS was used for evolutionary studies, it deviates from 1.373 by no more than 4.5%. This correction would explain the discrepancies noted above, particularly the larger ones reported by Jones et al (1998) and Ebeling et al (1999).

Once we have opened the Pandora box of trying to determine a value of f that is accurate to ~10% other effects must also be included. No cluster has X-ray emission at infinite radius, as we have assumed so far. In fact, the emission will only extend to the virial radius, given according to the theory outlined in Section 4 by, $r_v(z_v,kT) = (3M/[4\pi\Delta(\Omega_0,z_v)\ \Omega_0\ \rho_{c,0}(1 + z_v)^3])^{1/3} = 1.90\ h^{-1}$ Mpc $\beta_{TM}^{1/2}$ $(kT/10keV)^{1/2}$ $(1 + z_v)^{-3/2}$ (for $\Omega_0 = 1$) $= 2.08\ h^{-1}$ Mpc $(kT/10keV)^{1/2}$ $(1 + z_v)^{-3/2}$ (for $\beta_{TM}$ given in Section 4). Most of these terms are defined after equation (12). For now we note that $z_v$ is the virialization redshift that we take to be the redshift of the cluster. In addition, Vikhlinin, Forman, & Jones (1999) have shown that the surface brightness in the outer regions of clusters falls off faster than that given by our standard $\beta = 2/3$. This value was determined with Einstein Imaging Proportional Counter (IPC) measurements that were not as sensitive as the ROSAT Position Sensitive Proportional Counter (PSPC) that is able to trace the cluster emission out to larger radii. From the data in Vikhlinin et al (1999) we find $\beta(kT) = (0.621 \pm 0.022) + (0.017 \pm 0.004)\ kT$. Including the PSF, the new value of $\beta$ and retaining the fixed core radius given above, but only integrating out to the virial radius yields a new f, $f_{PSFkT}$, that is ~6% larger than f when averaged over the ranges $0.3 < z < 0.6$ and $4 < kT < 10$ appropriate to our situation. The range of $f_{PSFkT}$ over the same range of z and kT is 0.98 - 1.13. While a correction of this size would probably resolve the EMSS - ASCA flux discrepancy, it is likely too small for that noted by Jones et al (1998) and Ebeling et al (1999). And still we have not yet included the different core radii of different clusters. Furthermore, for our purposes, we only need the one-half power of the total flux to determine the volume accessible to a given temperature cluster (see Section 3.2), which gives only a few percent change from the standard EMSS f. Given the remaining uncertainties and the smallness of the changes after including everything described above, we have retained the original EMSS fraction given by equation (3).

Given the wide range of exposures on the targets in the EMSS, there is a correspondingly wide range of solid angles surveyed to a particular flux limit. There is not a single flux limit for the entire survey region as there is for the low redshift sample. The integral sky coverage, $\Omega_{surv}$, as a function of limiting flux, is in Table 3 of H92. We repeat it here in our Table 3 and also give the differential sky coverage, $d\Omega_{surv}$, that is the additional solid angle surveyed upon going to the next higher flux limit. The maximum redshift, $z_{max}$, to which a cluster at redshift z and detect cell flux $F_{det}$ can be detected for the $i^{th}$ flux limit, is given by equation (3) of H92:

$$F_{\lim,i}(z_{\max,i}) = F_{\det}\,\frac{D_L^2(1,z)}{D_L^2(1,z_{\max,i})}\,\frac{f(\theta_D/\theta_0(z_{\max,i}))}{f(\theta_D/\theta_0(z))} \tag{4}$$

where $D_L(\Omega_0,z)$ is the luminosity distance (see the Appendix). Note that in equations (3) and (4) above $\Omega_0 = 1$ is to be used because the value of $a_0$ was derived using that assumption. The volume searched for each cluster is then obtained by summing over all flux limits:

$$V_{sea} = \sum_i \left[ \frac{dV(\Omega_0, \le \min(z_u, z_{\max,i}))}{d\Omega} - \frac{dV(\Omega_0, \le z_l)}{d\Omega} \right] d\Omega_{surv,i} \tag{5}$$

where $z_l$, and $z_u$ are the lower and upper redshift boundaries respectively (0.3 and 0.6 for our case) and $dV(\Omega_0,<z)/d\Omega$ is the volume per unit solid angle within a redshift z given in the Appendix.

We give these volumes in Table 2 using the detect cell flux derived by multiplying the total flux from our ASCA observations by the f appropriate for each cluster. By using the ASCA data to determine the detect cell flux instead of the original EMSS data, we have decided to accept a ~15% systematic uncertainity instead of ~25% statistical uncertainity. The volumes were calculated assuming $\Omega_0 = 0.5$. The median search volume for clusters in the high redshift sample is 6.4 x $10^7$ and 7.7 x $10^7$ $h^{-3}$ $Mpc^3$ for a zero cosmological constant universe (termed open) and for a spatially flat universe in which $\Omega_0 + \lambda_0 = 1$ (termed flat) respectively. These volumes are comparable to that for the low redshift sample. We also give the integral temperature functions in Table 2 and in Figures 1 and 2 as calculated from equation (1).

## 3. MAXIMUM LIKELIHOOD METHOD AND TEMPERATURE SELECTION FUNCTION

### 3.1 Maximum Likelihood

We use a maximum likelihood fit to the unbinned data in order to determine various model parameters. The method is described in Marshall et al (1983). The likelihood function is given by their equation (2), which for our situation is:

$$S = -2\sum_{i=1}^{N} \ln\left[ n(\Omega_0, z_i, kT_i)\frac{d^2V(\Omega_0, z_i)}{dzd\Omega} \right] + 2\int_{kT\min}^{kT\max} dkT \int_{z\min}^{z\max} n(\Omega_0, z, kT)\Omega(z,kT)\frac{d^2V(\Omega_0,z)}{dzd\Omega}dz \tag{6}$$

Here N is the number of clusters observed, $n(\Omega_0,z,kT)$ is the temperature function, $\Omega(z,kT)$ is the solid angle in which a cluster with temperature kT at redshift z could have been detected (the selection function) and $d^2V/dzd\Omega$ is the differential volume, which is given in the Appendix. The $d^2V/dzd\Omega$ in the first term of equation (6) is not in Marshall et al (1983) because it is a function of one of the parameters ($\Omega_0$) we are trying to constrain. The usual procedure is to assume a value for this parameter. This term was not in the likelihood function of H97. Its inclusion has only a small effect; the best fit values with and without it agree within their 68% confidence errors. However, we retain it here.

The best estimates of the model parameters and derived quantities are obtained by minimizing S. Confidence regions for them are obtained by noting that S is distributed as $\chi^2$ with the number of degrees of freedom equal to the number of interesting parameters (Lampton, Margon and Bowyer, 1976; Avni 1976). We have either one, two or three interesting parameters depending on what we are trying to constrain. Sometimes we are only interested in $\Omega_0$, at other times we want to know the full temperature function or to derive constraints on $\Omega_0$ and $\sigma_8$ (three parameters).

The maximum likelihood fit only provides the best fit, it does not provide an assessment of whether that fit is a good fit to the data. Marshall et al (1983) require that the two independent cumulants

$$C(\leq z_i) = \frac{\int_{z\,\min}^{z_i} dz \int_{kT\,\min}^{kT\,\max} n(\Omega_0, z, kT)\Omega(z, kT)(d^2V(\Omega_0, z)/dzd\Omega)dkT}{\int_{z\,\min}^{z\,\max} dz \int_{kT\,\min}^{kT\,\max} n(\Omega_0, z, kT)\Omega(z, kT)(d^2V(\Omega_0, z)/dzd\Omega)dkT} \quad (7)$$

and

$$C(\geq kT_i \mid z_i) = \frac{\int_{kTi}^{kT\,\max} n(\Omega_0, z_i, kT)\Omega(z_i, kT)dkT}{\int_{kT\,\min}^{kT\,\max} n(\Omega_0, z_i, kT)\Omega(z_i, kT)dkT} \quad (8)$$

both be uniform according to the two-tailed Kolmogorov-Smirnov test. They recommend rejecting the model if the probability of either being uniform is less than 0.05.

### 3.2 Temperature-Luminosity Relation and Selection Function

The clusters in both the low and high redshift samples were selected from flux limited surveys. They were not selected from a temperature-limited survey; such a survey may not even be possible. Fortunately, clusters obey a temperature – luminosity relation and the luminosity may be derived from the redshift and flux. Thus we recast the selection as a function flux into a selection as a function of redshift and luminosity and, via the temperature – luminosity relation, as a function of redshift and temperature.

There have been many discussions of the cluster temperature – luminosity relation. Among the larger and/or more recent compilations are David et al. (1993, 104 clusters), White, Jones, and Forman (1997, 86 clusters), Mushotzky and Scharf (1997, 38 clusters), Markevitch (1998, 35 clusters), Allen and Fabian (1998, 30 clusters), Arnaud and Evrard (1999, 24 clusters) and Henry and Sornig (1999, 29 clusters). The relation from the data in Table 2 is shown in Figure 4. The conclusions from all this work are that clusters have a definite temperature – luminosity relation that exhibits little or no evolution at any redshift but the relation has a large scatter related to the size of the cooling flow at the cluster center. Not all of these compilations are applicable to our work here because some have included the effects of cooling flows in their analysis, either by excluding the central region, or by choosing samples that contain mostly cooling flow or non-cooling flow clusters. It is not yet generally possible to discern the existence of cooling flows at high redshift, much less correct for them.

We give in Table 4 a selection of recent results that are most relevant to this paper. The two low redshift samples analyze essentially all available data and use mostly the same clusters. The Henry and Sornig (1999) result comes from all publicly available ASCA observations of clusters as of July 1, 1998 with redshifts between 0.3 and 0.6 and contains the objects in this paper as well as others. None of the fits given differentiate between cooling flow and non-cooling flow clusters. It is apparent that there is little evolution between z = 0.39 and 0.06 and therefore we use the David et al (1993) fit to determine our selection functions for both samples.

There are two additional complications. First, in order approximately to compensate for the large scatter in the temperature – luminosity relation we have used the –1 σ values of both the normalization and slope of the David et al (1993) fit. Second, we convert the fit in Table 4 from the bolometric luminosity to the band in which the clusters were selected. Specifically, for the low redshift sample we use $kT = 3.55[L_{44}(2,10)]^{0.286}$ and for the high redshift sample we use $kT = 3.05[L_{44}(0.3,3.5)]^{0.355}$, after allowing for both effects. The selection functions are given in Figure 5, using f in equation (3) for the high redshift case. A comparison with the data from Tables 1 and 2 indicates they provide a plausible description of the selection. The most discrepant cluster is C0336, one of only two non-Abell clusters in the low redshift sample. This object is an extended X-ray source not contaminated by point sources (Irwin & Sarazin, 1995).

## 4. THEORETICAL TEMPERATURE FUNCTION

The primordial fluctuation spectrum is the squared modulus of the spatial Fourier transform of the fluctuations in the underlying matter density field about their average. We use a power law model for this quantity:

$$P(k) = \frac{1}{k_0^3 V_u}\left(\frac{k}{k_0}\right)^n \tag{9}$$

In this expression the shape parameter is n and $V_u$ is a large volume in which the statistical properties of the fluctuations are the same as in any other such volume. The normalization is set by $k_0$. Nearly always the normalization is parameterized with the rms mass fluctuations on an $8h^{-1}$ Mpc scale, $\sigma_8$. This power spectrum gives the rms value of the density fluctuations at zero redshift when smoothed with a top hat containing a mass M:

$$\sigma_\rho(\Omega_o, 0, M) = 0.675\left[\frac{\Gamma(3+n)\sin(n\pi/2)}{2^n n(2+n)(1-n)(3-n)}\right]^{1/2}\left[\left(9.5k_o h^{-1}Mpc\right)^3 \Omega_o^{-1} h M_{15}\right]^{-(3+n)/6} \tag{10}$$

(Peebles, 1980) where $M_{15}$ is the mass in units of $10^{15}$ solar masses. Then, $\sigma_8 = \sigma_\rho(\Omega_o, 0, 0.594 \times 10^{15}\, h^{-1}\Omega_o)$ which is independent of $\Omega_o$.

The spatial distribution of the fluctuations that grow by gravity into the structure we see is determined by P(k). The rate of their growth is determined mainly by $\Omega_0$. What objects can we use to trace this evolution? Probably the best are X-ray emitting clusters of galaxies. The physics and astrophysics of X-ray emission from clusters of galaxies are relatively straightforward. Simple gravitational processes dominate cluster formation and evolution and imply that clusters are still forming today. A cluster itself is non-linear, having collapsed, but it formed out of a background mass density field that is still evolving linearly, an enormous simplification. These same formation processes also heat gas trapped by the cluster potential, which then produces optically thin thermal radiation. The gas is in simple coronal equilibrium; that is the ion populations are mostly in the ground state and electron impact dominates the ionization balance and excitation processes.

Given the simplicity of the relevant processes, there is hope that the evolution of cluster X-ray emission can be reliably calculated and those calculations may be verified by direct observations of nearby objects. This situation is in stark contrast to that for supernovae, quasars and galaxies, the only other objects that are observable at cosmological distances. The evolution of these objects is too complicated to be reliably calculated and/or they formed too long ago for it to be well observed.

The evolution of the cluster mass function is given by the Press-Schechter (1974, PS) formula. Although relatively simple, PS has been repeatedly shown to provide a very accurate description of N-body calculations (e.g. Lacey and Cole, 1994; Mo and White, 1996), at least over the range of masses of interest to us (Gross et al, 1998). The number of objects per unit mass interval per comoving volume is, according to PS:

$$n(\Omega_o, z, M) = -\sqrt{\frac{2}{\pi}}\frac{\Omega_o \rho_{c,o}}{M}\frac{\delta_c(\Omega_o, z)}{\sigma_\rho(\Omega_o, 0, M)}\frac{D(\Omega_o, 0)}{D(\Omega_o, z)}\frac{d\ln\sigma_\rho(\Omega_o, 0, M)}{dM}$$

$$\exp\left[\frac{-\delta_c^2(\Omega_o, z)D^2(\Omega_o, 0)}{2\sigma_\rho^2(\Omega_o, 0, M)D^2(\Omega_o, z)}\right] \qquad (11)$$

where $\rho_{c,o}$ is the present critical density, $\delta_c(\Omega_o,z)$ is the mass density fluctuation required for a spherical perturbation to collapse at redshift z, and $D(\Omega_o,z)$ is the growth factor of perturbations while they are evolving linearly. Recent calculations have now derived these quantities for most cases of interest. We give them in the Appendix for the three model universes we consider here.

However, the cluster mass is not an easily measurable quantity, although it may be determined from X-ray observations in some circumstances. More readily observable are the X-ray luminosity or temperature. The X-ray luminosity depends on the square of the hot gas density, which may have its own evolution independent of that of the cluster. Perhaps the best cluster observable readily accessible to current instrumentation is the integrated temperature. As one might suspect, and numerical hydrodynamic calculations confirm the suspicion, the average cluster temperature is given by the depth of the potential well of the dark matter of the cluster. If the gas were hotter (cooler) than that given by virial equilibrium, then it would have blown away (collapsed) long ago because the time scale for changes in the configuration of the gas, the sound crossing time, is much shorter than the Hubble time.

The temperature function is obtained from the mass function of equation (11) via the chain rule; n(kT) = n(M) dM/dkT. The mass-temperature relation is given by considering a modified singular isothermal sphere:

$$kT = \frac{7.98 kev}{\beta_{TM}}\left[\frac{\Omega_o \Delta(\Omega_o, z_v)}{18\pi^2}\right]^{1/3}(hM_{15})^{2/3}(1+z_v)\,. \qquad (12)$$

Here $\Delta(\Omega_o,z_v)$ is the ratio of the cluster's average density to that of the background mass density at the virialization redshift (see the Appendix) and $\beta_{TM}$ is the modification factor accounting for departures from virial equilibrium. Bryan & Norman (1998, their Figure 5) show that equation (12) is a good description of the mass – temperature relation in numerical hydrodynamic simulations over at least the redshift range 0 – 1.

We determine $\beta_{TM}$ from numerical hydrodynamic simulations of clusters, the results from which we give in Table 5. The average value is $1.21 \pm 0.05$. In an intriguing paper on this issue, Shapiro, Iliev and Raga (1998) have derived a value of 1.16 by considering a truncated isothermal sphere that has minimum energy. Physically, the pressure to provide the truncation comes from the outward going shock front in a spherical infall solution. This new work provides the final piece of a completely analytic model of cluster evolution. In the future it may be desirable to determine $\beta_{TM}$ empirically, but at present such determinations have a quite substantial error as the last three rows in Table 5 show. For the moment, the empirical determinations do not disagree with those from the hydrodynamic simulations or the analytic result.

Combining equations (10), (11) and (12) gives the evolving cluster temperature function.

$$n(\Omega_0, z, kT) = 1.29x10^{-5}(h^{-1}Mpc)^{-3}(keV)^{-1}\Omega_0\delta_c(\Omega_0, z)\frac{D(\Omega_0,0)}{D(\Omega_0,z)}(1+z_v)^{(3-n)/4}(9.5h^{-1}Mpc\Omega_0^{-1/3}k_0)^{(3+n)/2}$$

$$\left(\frac{\beta_{TM}kT}{7.98}\right)^{(n-7)/4}\left(\frac{18\pi^2}{\Omega_0\Delta(\Omega_0 z_v)}\right)^{(n-3)/12}\left\{\frac{2^n n(1-n)(3-n)(2+n)(3+n)^2}{\Gamma(3+n)\sin(n\pi/2)}\right\}^{1/2}$$

$$\exp\left[-1.1\delta_c^2(\Omega_0, z)\frac{D^2(\Omega_0,0)}{D^2(\Omega_0,z)}(1+z_v)^{-(3+n)/2}\left(\frac{\beta_{TM}kT}{7.98}\right)^{(3+n)/2}\left(\frac{18\pi^2}{\Omega_0\Delta(\Omega_0, z_v)}\right)^{(3+n)/6}\right.$$

$$\left.(9.5h^{-1}Mpc\Omega_0^{-1/3}k_0)^{3+n}\left\{\frac{2^n n(1-n)(3-n)(2+n)}{\Gamma(3+n)\sin(n\pi/2)}\right\}\right] \quad (13)$$

We will also need the integral function, which is

$$n(\Omega_0, z, > kT) = 2.58x10^{-5}(h^{-1}Mpc)^{-3}(1.1)^{(3-n)/(2n+6)}\frac{7.98}{\beta_{TM}}\left(\delta_c(\Omega_0, z)\frac{D(\Omega_0,0)}{D(\Omega_0,z)}\right)^{6/(n+3)}(9.5h^{-1}Mpc\ k_0)^3$$

$$\left\{\frac{2^n n(1-n)(3-n)(2+n)}{\Gamma(3+n)\sin(n\pi/2)}\right\}^{3/(n+3)}\Gamma\left[\frac{n-3}{2(n+3)}, 1.1\left(\delta_c(\Omega_0, z)\frac{D(\Omega_0,0)}{D(\Omega_0,z)}\right)^2(1+z_v)^{-(n+3)/2}\right.$$

$$\left.\left(\frac{18\pi^2}{\Omega_0\Delta(\Omega_0, z_v)}\right)^{(3+n)/6}(9.5h^{-1}Mpc\Omega_0^{-1/3}k_0)^{3+n}\left\{\frac{2^n n(1-n)(3-n)(2+n)}{\Gamma(3+n)\sin(n\pi/2)}\right\}\left(\frac{\beta_{TM}kT}{7.98}\right)^{(n+3)/2}\right]. \quad (14)$$

We adopt the late collapse approximation and set the redshift at which the cluster is observed equal to the virialization redshift, $z = z_v$. It is possible to relax this approximation, but the effects of doing so are not large. Figure 8a of Kitayama & Suto (1996), Figure 5 of Kitayama & Suto (1997) and Figure 3 of Viana and Liddle (1999) compare various results with and without the approximation. Note that the parameter s = 0 in both papers by Kitayama & Suto corresponds to our equation (12). In general, there is almost no change at all for either the flat model (for redshifts under consideration here) or the open model for large $\Omega_0$. For example, the inferred value of $\Omega_0$ for the flat model changes by ~ 0.01 near $\Omega_0$ = 0.3 and z = 0.4, judging from Figure 3 of Viana and Liddle (1999). Only for $\Omega_0$ ~ 0.3 open models at z ~ 0.4 is the effect on $\Omega_0$ ~ 0.1, but this size is still comparable to the statistical uncertainity of the present samples.

## 5. MEASURING COSMOLOGICAL PARAMETERS

We are now ready to fit the data presented in Section 2 to the theory described in Section 4 using the maximum liklihood method of Section 3. The best fitting integral temperature functions are shown in Figures 1 and 2. The two cumulants given in equations (7) and (8) are uniform with a probability of 0.70 and 0.14 respectively for the open model and 0.78 and 0.13 respectively for the flat model. Thus each fit is acceptable.

The most general description of the results requires the three parameters of the fit. We show these results in Figures 6 and 7. It is straightforward to read off the value of n, which is $-(1.72^{+0.31}_{-0.37})$ and $-(1.68^{+0.35}_{-0.41})$ at the 68% confidence for the open and flat models respectively. These values agree well with that measured in HA91. The shape parameter that we adopt, n, may be converted to the shape parameter of the popular cold dark matter fluctuation spectrum, Γ, using equation (5) of Pen (1998) or equations (2) and (3) of Viana and Liddle (1999). Both give nearly the same results, which are that our fit implies $\Gamma \approx 0.05 \pm 0.22$.

The presentation in Figures 6 and 7 is somewhat difficult to appreciate, so we also give the constraints for fewer parameters. First we concentrate just on $\Omega_0$ in Figure 8. Constraints are relatively tight when considering this single parameter. We find that $\Omega_0 = 0.49\ ^{+0.13}_{-0.10}$ at the 68% confidence level and $^{+0.26}_{-0.19}$ at the 95% confidence level for the open model. For the flat model the corresponding values are $\Omega_0 = 0.44\ ^{+0.12}_{-0.11}$ at the 68% confidence level and $^{+0.26}_{-0.20}$ at the 95% confidence level. These results agree with those in H97, except the errors are ~ 40% smaller here. A closed universe is ruled out at much greater than 99% confidence.

We recast the constraints in Figures 6 and 7 into a more conventional format in Figures 9 and 10. Three parameters are still required, but the constraints on n and $k_0$ are collapsed into $\sigma_8$. Some of the degeneracy between these two parameters still remains in the form of a banana shaped contour particularly for the flat model. Still, we are able to make a relatively precise measurement of $\sigma_8$, finding values of $0.72^{+0.11}_{-0.08}$ and $0.77^{+0.20}_{-0.10}$ at 68% confidence level for the open and flat cases respectively. These results also agree with those in H97.

## 6. COMPARISON WITH OTHER MEASUREMENTS

Determining the values of cosmological parameters is one of the fundamental goals of astrophysics. Recently, the number of results has increased dramatically and it is impossible to provide a comprehensive review in one section of a paper reporting still more. We have given in the Introduction a short summary of what was obtained using clusters of galaxies utilizing similar methods to ours. Here we want to give a few of the most recent results using methods that are very different.

Our analysis in terms of the fluctuation spectrum, P(k), permits a direct measurement of this quantity at spatial frequencies approximately 0.2 to 0.5 h $Mpc^{-1}$. The positions of galaxies, either as a two dimensional angular distribution or a three dimensional spatial distribution, give a direct measurement of the shape (assuming the bias between galaxies and mass is independent of scale, eg Mann, Peacock, & Heavens, 1998), but not the normalization, of P(k) over a much wider range. Gaztañaga & Baugh (1998) have measured the local slope of P(k) using the angular correlation function of the APM Galaxy Survey. Galaxies are in the non-linear regime at the spatial frequencies probed by clusters, but after correcting for this effect the APM measurements imply that n = -(1.6 ± 0.2), which is in excellent agreement with our result.

The baryon fraction contained in clusters, combined with $\Omega_b$ provided by Big Bang nucleosynthesis, has been used to measure the same three cosmological parameters that we have. Evrard (1997) finds $\Omega_0 = (0.30 \pm 0.07)\ h^{-2/3}$, which is $\Omega_0 = 0.42 \pm 0.10$ for h = 0.6. Shimasaku (1997) determined the cluster gas mass function from which he derived $(\sigma_8, n) = (0.64^{+0.18}_{-0.09}, -1.6^{+1.1}_{-0.4})$ for h = 0.5 and $(\sigma_8, n) = (0.73^{+0.27}_{-0.11}, -1.5^{+1.2}_{-0.4})$ for h = 0.8. All of these results agree well with ours.

Deviations from a smooth Hubble flow, termed redshift distortions or peculiar velocities, caused by the gravity of mass fluctuations constrain the quantity $\Omega_0^{0.6}\sigma_8/\sigma_8(tracer)$, where $\sigma_8(tracer)$ is defined analogously to $\sigma_8$ using the objects tracing the mass field and velocities in it. The degeneracy between the parameters measured by this method is similar to that from cluster abundances at a single redshift. The peculiar velocity constraint has been applied mostly to samples of galaxies. Three recent analysis of the IRAS 1.2 Jy sample, where $\sigma_8$(IRAS) = 0.69 ± 0.04 (Fisher et al, 1994), are described by Sigad et al (1998), da Costa et al (1998), and Willeck et al (1998). They find $\Omega_0^{0.6}\sigma_8$ = 0.61 ± 0.09, 0.41 ± 0.07 and 0.34 ± 0.05 respectively. The discrepancy among these three analyses of nearly the same data is somewhat unexpected. A similar measurement using clusters of galaxies gives $\Omega_0^{0.6}\sigma_8 = 0.44^{+0.19}_{-0.13}$ (90% confidence, Watkins, 1997; Borgani et al, 1997). We compare the da Costa et al (1998) result with ours in Figures 11 and 12.

The statistics of the number of gravitationally lensed quasars can constrain the cosmological constant. Some recent work and its 95% confidence limit (unless otherwise noted) on $\Omega_0$, assuming the flat model, is by Falco, Kochanek & Muñoz (1998; > 0.26), Cooray (1999; > 0.21), Cooray, Quashnock & Miller (1999; > 0.21), Helbig et al (1999; > 0.36), Chiba & Yoshii (1999; $0.3^{+0.2}_{-0.1}$ at 68% confidence) and Cheng & Krauss (1999; 0.25 – 0.55 systematic effects dominate). All these results agree with what we find.

Fluctuations in the cosmic microwave background (CMB) determined with COBE normalize P(k). This normalization may be extrapolated from the scales probed by COBE to cluster scales if the shape of P(k) is known. Bunn & White (1997), among many others, use a CDM spectrum to perform the extrapolation. Their results are shown in Figures 11 and 12. The CMB data nicely complement that from clusters and both combine to yield a small overlap region for either the open or flat models that results in a somewhat lower value of $\Omega_0$ than our best fit.

A classical test used to determine cosmological parameters is the Hubble diagram. Collins and Mann (1998) present a K-band Hubble diagram for brightest cluster galaxies in X-ray luminous clusters that has very small scatter. They find good agreement with us, with $\Omega_0$ = 0.28 ± 0.24 for the open model and $0.55^{+0.14}_{-0.15}$ for the flat model (both at 68% confidence).

Of course one of the most productive uses of the Hubble diagram recently has come from the one constructed from Type Ia supernovae. Efstathiou et al (1999) combine this diagram with measurements of fluctuations in the CMB at COBE and smaller scales. They exclude the open model at high confidence but the flat model is quite consistent with their data. If we limit their results to such a model, in order to compare with our own, they find $\Omega_0$ = 0.29 ± 0.08 (at 68% confidence). Tegmark (1999) has emphasized that including additional physics, such as gravity waves, will substantially increase the size of the error bars. Nevertheless, these results are consistent with ours.

Although the above has not been comprehensive, we hope it has been representative. Most recent analyses indicate that $\Omega_0$ lies approximately in the range 0.25 - 0.55. As useful as the comparison of results among different methods is, a better approach is to combine several data sets in order to determine a joint best fit. This approach has at least three advantages: smaller errors, breaking of degeneracies among the parameters, and possible revelation of systematic errors. Webster et al (1998), Lineweaver (1999), and Bridle et al (1999) have taken initial steps in this direction. The latter work performs a simultaneous fit to the CMB fluctuations, the IRAS 1.2 Jy redshift survey and the data presented here. There is good agreement among all three data sets and with our results.

## 7. FUTURE WORK

All but one EMSS cluster with z > 0.3 has now been observed with ASCA. The last one is unobservable with ASCA due to its proximity to NGC4151, which is very bright. Thus it may be possible to repeat the analysis here with ~ 20 high redshift clusters, a number comparable to the present low redshift sample. In addition to improving the statistics because of increased sample size, some of the additional clusters will be at higher redshifts than any used here and that provides increased leverage on the measurement.

Another statistical error is that on the temperature measurement. Since the temperature function is a very steeply falling function, simple statistical errors on temperature bias it high. Further, the errors on temperature are larger for the high redshift sample compared to the low, which may lead to some bias. We are probably close to the Time Allocation Committee limit, so it is not feasible to improve greatly on this situation with ASCA. However, the upcoming XMM mission will provide a large increase in effective area that should substantially improve the statistical quality of the temperatures.

Useful as an improvement in statistics will be, we believe that the present measurement is starting to be limited by systematic uncertainties. We have already discussed in Section 4 the effects of assuming all clusters collapse at the epoch they are observed. Another systematic effect comes from the cluster temperature - luminosity relation, which is integral to the determination of the selection function. This relation has a very large scatter, but the scatter is reduced considerably if the effects of the central cooling flow are removed (Allen & Fabian, 1998; Markevitch, 1998). It is only possible to make this exclusion with ASCA for clusters at low redshift, unless the observation has a sufficiently large number of detected photons that multiple component spectral models can be fit. As mentioned before, the ASCA exposures are already quite long so this approach is not possible for us. However, both the Chandra and upcoming XMM missions will permit a simple spatial exclusion of the cooling flow. Robinson, Gawiser, & Silk (1999) and Eke et al (1998) suggest another method of limiting the systematic effects caused by the temperature – luminosity relation. The hottest clusters in a redshift limited sample are detectable in the entire volume defined by the redshift cuts because they are so luminous. Thus the temperature – luminosity relation is mostly irrelvant. Viana & Liddle (1999) also discussed the utility of working only with the hottest clusters.Unfortunately, adopting this approach with the present samples yields very large statistical errors because there are so few clusters hot enough.

We believe the best way to eliminate the effects of systematic errors is to repeat the analysis performed here with completely new samples. Several such samples, both at low and high redshift, have now been compiled from ROSAT. A relatively modest investment of XMM time, comparable to what has already been used here with ASCA, would permit a totally independent measurement of the same cosmological constants to better statistical precision.

Notwithstanding the desirable additional work described above, we want to end on a positive note. Most independent methods imply that the universe will expand forever. We have an empirical determination of the mass fluctuation spectrum. Clusters abundances and CMB fluctuations provide the normalization at the small and large scales and galaxy redshift surveys provide the shape in between. The combination of these two tells the story of how the magnificent large scale structures we see about us came to be.

I am pleased to acknowledge many useful discussions with Harald Ebeling and Vincent Eke, which greatly improved my understanding of the maximum likelihood technique, the EMSS selection function and the theory of cluster evolution. E. Miyata is the Co-PI of the MS1621.5 and MS2053.7 observations. I also have benefited from conversations with K. Arnaud, M. Arnaud, N. Bahcall, A. Blanchard, S. Cole, S. De Grandi, M. Donahue, A. Evrard, C. Frenk, I. Gioia, M. Hattori, J. Irwin, A. Liddle, C. Mullis, J. Navarro, M. Norman, J. Ostriker, J. Primack, R. Sadat, P. Shapiro, Y. Suto, M. Turner, P. Viana, A.Vikhlinin, M. Voit, and S. White. Over the years this work has been supported by NASA through grants NAG5-2523 and NAG5-4828.

## APPENDIX

Here we collect the various quantities, defined in the body of the paper, that are required for our theoretical analysis of the temperature function and maximum likelihood fits to our data. As a reminder, $\delta_c(\Omega_o,z)$ is the critical overdensity for collapse at redshift z; $\Delta(\Omega_o,z)$ is the ratio of the average cluster mass density to that of the background density for a cluster which collapses at redshift z; $D(\Omega_o,z)$ is the growth factor up to redshift z; $D_L(\Omega_o,z)$ is the luminosity distance to redshift z and the angular diameter distance is $D_L(\Omega_o,z)/(1+z)^2$; V is the volume and $d\Omega$ is the solid angle surveyed. Quantities with subscript zero are evaluated at the present epoch. We distinguish three cosmologies, closed, open and flat.

Closed Model $\Omega_o = 1$

$$\delta_c(1,z) = \frac{3(12\pi)^{2/3}}{20} \approx 1.686$$

(Peebles, 1980, Eq. 19.48a evaluated at $2\pi$)

$$\Delta(1,z) = 18\pi^2 \approx 177.7$$

(Peebles, 1980, Eq. 19.50 multiplied by 32. A factor of 8 results from assuming the cluster collapses to a radius half its value at maximum expansion and another factor of 4 from the decreasing background density during the time required to collapse that is twice the time to reach maximum radius.)

$$D(1,z) = \frac{1}{1+z}$$

(Peebles, 1980, Eqs. 11.5,11.7 and 92.2)

$$D_L(1,z) = 2\frac{c}{H_0}(1+z-\sqrt{1+z})$$

$$\frac{dV(1,\le z)}{d\Omega} = \frac{8}{3}\left(\frac{c}{H_0}\right)^3\left(1-\frac{1}{\sqrt{1+z}}\right)^3$$

$$\frac{d^2V(1,z)}{dzd\Omega} = 4\left(\frac{c}{H_0}\right)^3\frac{2+z-2\sqrt{1+z}}{(1+z)^{5/2}}$$

Open Model $\Omega_o < 1$, $\lambda_o = 0$

$$x \equiv 2\frac{\Omega_0^{-1}-1}{1+z}$$

$$\Omega(z) = \frac{2}{2+x}$$

$$\delta_c(\Omega_0, z) = \frac{3}{2}\left(1 + (2\pi)^{2/3}\left\{\sqrt{x^2+2x} - \ln\left[(1+x) + \sqrt{x^2+2x}\right]\right\}^{-2/3}\right)\left(1 + \frac{6}{x} - \frac{3\sqrt{2+x}\ln\left[(1+x)+\sqrt{x^2+2x}\right]}{x^{3/2}}\right)$$

$$\Delta(\Omega_0, z) = \frac{(2\pi)^2 x^3}{\left\{\sqrt{x^2+2x} - \ln\left[(1+x)+\sqrt{x^2+2x}\right]\right\}^2}$$

(Maoz, 1990 Eq 12; Lacey and Cole, 1993 Eq A15; Oukbir and Blanchard, 1997)

$$D(\Omega_0, z) = \frac{5}{2x_0}\left\{1 + \frac{6}{x} - \frac{3\sqrt{2+x}\ln\left[(1+x)+\sqrt{x^2+2x}\right]}{x^{3/2}}\right\}$$

(Peebles, 1980 Eq. 11.16; our x is twice his; the different normalizing factor yields the same form as $\Omega_o \to 1$ as the closed model.)

$$D_L(\Omega_0, z) = \frac{2}{\Omega_0^2}\frac{c}{H_0}\left[\Omega_0 z + (2-\Omega_0)(1 - \sqrt{1+\Omega_0 z})\right]$$

$$\frac{dV(\Omega_0, \le z)}{d\Omega} = \left(\frac{c}{H_0}\right)^3 \left\{\frac{1}{2(1-\Omega_0)^{3/2}}\left[\ln\left(\frac{\sqrt{1+\Omega_0 z} + \sqrt{1-\Omega_0}}{\sqrt{1+\Omega_0 z} - \sqrt{1-\Omega_0}}\right) - \ln\left(\frac{1+\sqrt{1-\Omega_0}}{1-\sqrt{1-\Omega_0}}\right)\right]\right.$$
$$+ \frac{2-\Omega_0}{\Omega_0^2(1-\Omega_0)} - \frac{(2-\Omega_0+\Omega_0 z)\sqrt{1+\Omega_0 z}}{\Omega_0^2(1-\Omega_0)(1+z)^2}$$
$$\left. + \frac{8\sqrt{1+\Omega_0 z}}{\Omega_0^4(1+z)^2}\left[(2-\Omega_0)(\sqrt{1+\Omega_0 z} - 1) - \Omega_0 z\right]\right\}$$

This equation is numerically unstable at $\Omega_o = 1$ and for low $\Omega_o$ and/or z. Consequently, it is more reliable to integrate the following equation over z.

$$\frac{d^2V(\Omega_0, z)}{dz d\Omega} = 4\left(\frac{c}{H_0}\right)^3 \frac{\left[\Omega_0 z + (2-\Omega_0)(1-\sqrt{1+\Omega_0 z})\right]^2}{\Omega_0^4(1+z)^3\sqrt{1+\Omega_0 z}}$$

Flat Model $\Omega_o + \lambda_o = 1$

$$x \equiv \frac{\left(\Omega_0^{-1} - 1\right)^{1/3}}{1+z}$$

$$\Omega(z) = \frac{1}{1+x^3}$$

$$\delta_c(\Omega_0, z) = \frac{3(12\pi)^{2/3}}{20}\left[1 - 0.0123\log(1 + x^3)\right]$$ (Nakamura and Suto, 1997, Eq. C-28; Eke et al, 1996)

$$\Delta(\Omega_0, z) = 18\pi^2(1 + 0.4093x^{2.71572})$$ (Nakamura and Suto, 1997, Eq. C-19 their w is our $x^3$; Eke et al, 1996)

$$D(\Omega_0, z) = \frac{x}{x_0}\sqrt{1 + x^3}\int_0^1 dy(1 + x^3 y^{6/5})^{-3/2}$$ (Peebles, 1984, Eq. 14)

$$D(\Omega_0, z) = \frac{5 \cdot 3^{1/4}\sqrt{\Omega_0(1+z)^3 + (1-\Omega_0)}}{3x_0\sqrt{1-\Omega_0}}\left[E\left(\cos^{-1}\left[\frac{1+(1-\sqrt{3})x}{1+(1+\sqrt{3})x}\right], \kappa\right) - \frac{1}{3+\sqrt{3}}F\left(\cos^{-1}\left[\frac{1+(1-\sqrt{3})x}{1+(1+\sqrt{3})x}\right], \kappa\right)\right] + \frac{5(x^2 - (1+\sqrt{3}))}{3x_0(x + (1+\sqrt{3})x^2)}$$ (Einsenstein, 1998, Eqs. 8 - 10)

where $\kappa = \sin 75^0 = 0.5\sqrt{2+\sqrt{3}}$ . Here F(φ,κ) and E(φ,κ) are incomplete Legendre elliptic integrals of the first and second kind. Sometimes the following recursion relation must be used when φ > π/2: F(φ,κ) = 2 F(π/2,κ) - F(π - φ,κ) (Abramowitz and Stegun, 1964, equations 17.3.2 and 17.4.3) and similarly for E.

$$D_L(\Omega_0, z) = \frac{c}{H_0}(1+z)\int_0^z \frac{dy}{\sqrt{\Omega_0(1+y)^3 + (1-\Omega_0)}}$$ (Yoshii, 1995, Eq. 12; Caroll, Press, & Turner, 1992, Eqs. 23 and 25)

$$\frac{dV(\Omega_0, \le z)}{d\Omega} = \left(\frac{c}{H_0}\right)^3 \frac{1}{3}\left\{\int_0^z \frac{dy}{\sqrt{\Omega_0(1+y)^3 + (1-\Omega_0)}}\right\}^3$$ (Caroll, Press, & Turner, 1992, Eq. 26)

$$\frac{d^2V(\Omega_0, z)}{dz d\Omega} = \left(\frac{c}{H_0}\right)^3 \frac{1}{\sqrt{\Omega_0(1+z)^3 + (1-\Omega_0)}}\left\{\int_0^z \frac{dy}{\sqrt{\Omega_0(1+y)^3 + (1-\Omega_0)}}\right\}^2$$ (Yoshii, 1995, Eq. 14)

The integral in the last three equations is

$$\frac{1}{\Omega_0^{1/3}(1-\Omega_0)^{1/6}3^{1/4}}\left\{F\left(\cos^{-1}\left[\frac{1+(1-\sqrt{3})x_0}{1+(1+\sqrt{3})x_0}\right], \kappa\right) - F\left(\cos^{-1}\left[\frac{1+(1-\sqrt{3})x}{1+(1+\sqrt{3})x}\right], \kappa\right)\right\}$$

(see Mathiesen and Evrard, 1998 and references in Eisenstein, 1998). The numerical evaluation of the integrals is probably somewhat simpler to code than the Legendre closed form. Feige (1992) also gives an extensive discussion of the use of elliptical integrals in cosmological formulas.

We graph the critical overdensity for collapse in Figure 13. Note that this parameter is never more than a few percent different from its closed value (Lilje, 1992). We show in Figure 14 the growth factor to redshift z relative to the present epoch. It is only this combination that enters into the Press-Schechter mass function. Here, different values of the density parameter result in a large difference in growth. This difference is amplified by the exponential term of the Press-Schechter mass function. Finally, we show the cosmology dependent factors of the mass-temperature relation in Figure 15.

## REFERENCES

Abramowitz, M., & Stegun, I. A. 1964, Handbook of Mathematical Functions (Wahington, DC: US Government Printing Office)

Allen, S. W., & Fabian, A. C. 1998, MNRAS, 297, L57

Arnaud, K. A. 1996, Astronomical Data Analysis Software and Systems V, eds. Jacoby, G., & Barnes, J., p17, ASP Conf. Series vol. 101

Arnaud, M. 1994, in Cosmological Aspects of X-ray Clusters of Galaxies, W. C. Seitter ed., NATO ASI Series, 441, 197

Arnaud, M., & Evrard, A. E. 1999, MNRAS, 305, 631

Avni, Y. 1976, ApJ, 210, 642

Bahcall, N. A., & Cen, R. 1992, ApJ, 398, L81

Bahcall, N. A., Fan, X., & Cen, R. 1997, ApJ, 485, L53

Borgani, S., da Costa, L. N., Freudling, W., Giovanelli, R., Haynes, M. P., Salzer, J. & Wegner, G. 1997, ApJ, 482, L121

Borgani, S., Rosati, P., Tozzi, P., & Norman, C. 1999, ApJ, 517, 40

Bridle, S. L., Eke, V. R., Lahav, O., Lasenby, A. N., Hobson, M. P., Cole, S., Frenk, C. S., & Henry, J. P. 1999, MNRAS, in press

Bryan, G. L. & Norman, M. L. 1998, ApJ, 495, 80, 1998

Bunn, E. F. & White, M. 1997, ApJ, 480, 6, 1997

Carlberg, R. G., Yee, H. K. C., Ellingson, E., Abraham, R., Gravel, P., Morris, S., & Pritchet, C. J. 1996, ApJ, 462, 32

Carlberg, R. G., Morris, S. L., Yee, H. K. C., & Ellingson, E. 1997, ApJ, 479, L19

Carroll, S. M., Press, W. H., & turner, E. L. 1992, ARAA, 30, 499

Cheng, Y-C. N., & Krauss, L. M. 1999, astro-ph/9810393

Chiba, M., & Yoshii, Y. 1999, ApJ, 510, 42

Cole, S., & Kaiser, N. 1989, MNRAS, 237, 1127

Collins, C. A., & Mann, R. G. 1998, MNRAS, 297, 128

Cooray, A. R. 1999, AA, 342, 353

Cooray, A. R., Quashnock, J. M. & Miller, M. C. 1999, ApJ, 511, 562

da Costa, L. N., Nusser, A., Freudling, W., Giovanelli, R., Haynes, M. P., Salzer, J. J. & Wegner, G. 1997, MNRAS, 299, 425

David, L. P., Slyz, A., Jones, C., Forman, W., Vrtilek, S. D., & Arnaud, K. A. 1993, ApJ, 412, 479

Day, C. S. R., Fabian, A. C., Edge, A. C., & Raychaudhury, S. 1991, MNRAS, 252, 394

Donahue, M. & Voit, G. M., 1999 ApJ, 523, L137

Ebeling, H., Jones, L. R., Perlman, E., Scharf, C., Horner, D., Wegner, G., Malkan, M., Fairley, B., & Mullis, C. R. 1999, astro-ph/9905321

Edge, A. C., Stewart, G. C., Fabian, A. C., & Arnaud, K. A. 1990, MNRAS, 245, 559

Edge, A. C., & Stewart, G. C. 1991, MNRAS, 252, 414

Edge, A. C., Stewart, G. C., & Fabian, A. C. 1992, MNRAS, 258, 177

Efstathiou, G. Briddle, S. L., Lasenby, A. N., Hobson, M. P., & Ellis, R. S. 1999, MNRAS, 303, L47

Eke, V. R., Cole, S., & Frenk, C. S., 1996, MNRAS, 282, 263

Eke, V. R., Cole, S., Frenk, C. S. & Henry J. P. 1998 MNRAS, 298, 1145

Eke, V. R., Navarro, J. F., & Frenk, C. S. 1998, ApJ, 503, 569

Eisenstein, D. J. 1998, astro-ph/9709054

Evrard, A. E., 1989, ApJ, 341, L71

Evrard, A. E. 1990, in Clusters of Galaxies, ed. M. Fitchett, W. Oegerle, & L. Danley (Cambridge: Cambridge Univ. Press), 287

Evrard, A. E., 1997, MNRAS, 292, 289

Evrard, A. E., Metzler, C. A., & Navarro, J. F. 1996, ApJ, 469, 494

Falco, E. E., Kochanek, C. S., & Muñoz, J. A. 1998, ApJ, 494, 47

Fan, X., Bahcall, N. A., & Cen, R. 1997, ApJ 490, L123
Federspiel, M., Tammann, G. A., & Sandage, A. 1998, ApJ, 495, 115
Feige, B. 1992, Astron. Nachr. 313, 139
Fisher, K. B., Davis, M., Strauss, M. A., Yahil, A., & Huchra, J. P. 1994, MNRAS, 266, 50
Frenk, C. S., White, S. D. M., Efstathiou, G., & Davis, M. 1990, ApJ, 351, 10
Gaztañaga E., & Baugh C. M. 1998, MNRAS, 294, 229
Gioia, I. M., Maccacaro, T., Schild, R. E., Wolter, A., Stocke, J. T., Morris, S. L., & Henry, J. P. 1990, ApJS, 72, 567
Gioia, I. M., & Luppino, G. A. 1994, ApJ Supp 94, 583
Gross. M. A. K., Somerville, R. S., Primack, J. R., Holtzman, J., & Klypin, A. 1998, MNRAS, 301, 81
Hatsukade, I. 1989, ISAS Research Note 435
den Hartog, R., & Katgert, P. 1996, MNRAS, 279, 349
Henry, J. P., Soltan, A., Briel, U. G., & Gunn, J. E. 1982, ApJ, 262, 1
Henry, J. P., & Arnaud, K. A. 1991, ApJ, 372, 410 (HA91)
Henry, J. P., Gioia, I. M., Maccacaro, T., Morris, S. L., Stocke, J. T., & Wolter, A. 1992, ApJ, 386, 408
Henry, J. P. 1997, ApJ, 489, L1 (H97)
Henry, J. P., & Sornig, M. 1999, in preparation.
Helbig, P. et al. 1999, AA Suppl, 136, 297
Hjorth, J., Oukbir, J., & van Kampen, E. 1998, MNRAS, 298, L1
Horner, D. J., Mushotzky, R. F., & Scharf, C. A. 1999, astro-ph/9902151
Hughes, J. P., Butcher, J. A., Stewart, G. C., & Tanaka, Y. 1993, ApJ, 404, 611
Irwin, J. A. & Sarazin, C. L. 1995, ApJ, 455, 497
Jones, L. R., Scharf, C., Ebeling, H., Perlman, E., Wegner, G., Malkan, M., & Horner, D. 1998, ApJ, 495, 100
Kay, S. T. & Bower, R. G. 1999, MNRAS, 308, 664
Kitayama, T. & Suto, Y. 1996, ApJ, 469, 480
Kitayama, T. & Suto, Y. 1997, ApJ, 490, 557
Lacy, C., & Cole, S. 1993, MNRAS, 262, 627
Lampton, M., Margon, B., & Bowyer, S. 1976, ApJ, 208, 177
Lilje, P. 1992, ApJ, 386, L33
Lineweaver, C. H. 1999, Science, 284, 1503
McNamara, B. R., Bregman, J. N., & O'Connell, R. W. 1990, ApJ, 360, 20
Mann, R. G., Peacock, J. A. & Heavens, A. F. 1998, MNRAS, 293, 209
Maoz, D. 1990, ApJ, 359, 257
Markevitch, M., Forman, W. R., Sarazin, C. L., & Vikhlinin, A. 1998, ApJ, 503, 77
Markevitch, M. 1998, ApJ, 504, 27
Marshall, H. L., Avni, Y., Tananbaum, H., & Zamorani, G. 1983, ApJ, 269, 35
Mathiesen, B., & Evrard, A. E. 1998, MNRAS, 295, 769
Melnick, J., & Moles, M. 1987, Rev. Mex. Astron. Astrofis., 14, 72
Mo, H. J., & White, S. D. M. 1996, MNRAS, 282, 347
Molikawa, K., Hattori, M., Kneib, J-P., & Yamashita, K. 1998, preprint
Mohr, J. J., Mathiesen, B., & Evrard, A. E. 1999, ApJ, in press
Mushotzky, R. F., & Scharf, C. A. 1997, ApJ, 482, L13
Nakamura, T. T., & Suto, Y. 1997, Prog. Theor. Phys., 97, 49
Nichol, R. C., Holden, B. P., Romer, A. K., Ulmer, M. P., Burke, D. J., & Collins, C. A. 1997, ApJ, 644
Oukbir, J., & Blanchard, A. 1997, A&A, 262, L21
Oukbir, J., & Blanchard, A. 1997, A&A, 317, 1
Peebles, P. J. E. 1980, The Large-Scale Structure of the Universe (Princeton: Princeton Univ. Press)
Peebles, P. J. E. 1984, ApJ, 284, 439
Peebles, P. J. E., Daly, R., & Juszkiewicz, R. 1989, ApJ, 347, 563
Pen, U-L. 1998, ApJ, 498, 60

Perrenod, S. C. 1980, ApJ, 236, 373, 1980
Perrenod, S. C., & Henry, J. P. 1981, ApJ, 247, L1, 1981
Piro, L., & Fusco-Femiano, R. 1988, A&A, 205, 26
Press, W. H., & Schechter, P. 1974, ApJ, 187, 425 (PS)
Reichart, D. E. et al. 1999, ApJ, 518, 521
Robinson, J., Gawiser, E., & Silk, J. 1999, astro-ph/9906156
Sadat, R., Blanchard, A., & Oukbir, J. 1998, A&A, 329, 21
Shapiro, P. R., Illiev, I., & Raga, A. C. 1998, astro-ph/9810164
Shimasaku, K. 1997, ApJ, 489, 501
Sigad, Y., Eldar, A., Dekel, A., Strauss, M. A. & Yahil, A. 1998, ApJ, 495, 516
Struble, M. F., & Rood, H. J. 1987, ApJS, 63, 543
Tsuru, T. 1993, ISAS Research Note 528
Tamura, T., Day, C. S., Fukazawa, Y., Hatsukade, I., Ikebe, Y., Makishima, K., Mushotzky, R. F., Ohashi, T., Takenaka, K., Yamashita, K. 1996, PASJ, 48, 671
Tegmark, M. 1999, ApJ, 514, L69
Viana, P. T. P., & Liddle, A. R. 1996, MNRAS, 281, 323
Viana, P. T. P., & Liddle, A. R. 1999, MNRAS, 303,535
Vikhlinin, A., McNamara, B. R., Forman, W., Jones, C., Quintana, H. & Hornstrup, A. 1998, ApJ, 498, L21
Vikhlinin, A., Forman, W., & Jones, C. 1999, astro-ph/9905200
Watkins, R. 1997, MNRAS, 292, L59
Webster, A. M., Bridle, S. L., Hobson, M. P., Lasenby, A. N., Lahav, O., & Rocha, G. 1998, ApJ, 509, L65
White, S. D. M., & Rees, M. J. 1978, MNRAS, 183, 341
White, S. D. M., Efstathiou, G., & Frenk, C. S. 1993, MNRAS, 262, 1023
White, D. A., Jones, C., & Forman, W. 1997, MNRAS, 292, 419
Willick, J. A., Strauss, M. A.., Dekel, A. & Kolatt, T. 1997, ApJ, 486, 629
Yamashita, K. 1992, in Frontiers of X-ray Astronomy, Y. Tanaka & K. Koyama eds, (Universal Academy Press: Tokyo), p 475
Yoshii, Y. 1995, ApJ, 403, 552
Zabludoff, A. I., Huchra, J. P., & Geller, M. J. 1990, ApJS, 74, 1

FIGURE CAPTIONS

Figure 1. Integral cluster temperature function with best fitting open model superposed. The open (filled) circles are the low (high) redshift cluster observations, while the dotted (solid) lines are the low (high) redshift best-fitting models, respectively.

Figure 2. Integral cluster temperature function with best fitting flat model superposed. The open (filled) circles are the low (high) redshift cluster observations, while the dotted (solid) lines are the low (high) redshift best-fitting models, respectively.

Figure 3. Ratio of the ASCA to EMSS total fluxes as a function of the total flux measured with ASCA. The open circles are for clusters with no obvious point sources in their HRI images within 6.125' of the cluster center.

Figure 4. The data with $\pm$ 1$\sigma$ errors are the luminosity-temperature relation for clusters in the redshift range 0.3 – 0.6 reported here. The X-ray luminosity is in the 2-10 keV band and is calculated using h = 0.5 and $\Omega_0$ = 1 to be consistent with previous low-redshift work. The two solid lines are the $\pm$ 1$\sigma$ fits reported by David et al. (1993) for their low-redshift sample without any shift in normalization.

Figure 5. The selection functions for the low- and high-redshift cluster samples in the redshift-temperature plane with the data from Tables 1 and 2 superposed. The dotted lines on the high redshift selection function correspond to the EMSS integral sky coverage in Table 3, with the smallest solid angle at the coolest temperature. The solid angle for the low redshift selection function is a constant 8.23 steradians.

Figure 6. Sixty-eight percent confidence contours for the three interesting parameters n, $k_0$ and $\Omega_0$, corresponding to an increase in liklihood of 3.5 from the minimum, for the open model. Also shown are lines of constant $\sigma_8$.

Figure 7. Sixty-eight percent confidence contours for the three interesting parameters n, $k_0$ and $\Omega_0$, corresponding to an increase in liklihood of 3.5 from the minimum, for the flat model. Also shown are lines of constant $\sigma_8$.

Figure 8. Delta liklihood contours as a function of the single interesting parameter $\Omega_0$. The dotted lines show various confidence levels.

Figure 9. Sixty-eight percent confidence contour for $\sigma_8$ and $\Omega_0$ for the open model. The constraints on n and $k_0$ have been collapsed into $\sigma_8$, so three free parameters must still be considered. The contour thus corresponds to an increase in liklihood of 3.5 from the minimum.

Figure 10. Sixty-eight percent confidence contour for $\sigma_8$ and $\Omega_0$ for the flat model. The constraints on n and $k_0$ have been collapsed into $\sigma_8$, so three free parameters must still be considered. The contour thus corresponds to an increase in liklihood of 3.5 from the minimum.

Figure 11. The comparison of the constraints provided by the data presented in this paper with those from galaxy redshift distortions and the COBE normalization of a cold dark matter fluctuation spectrum for the open model.

Figure 12. The comparison of the constraints provided by the data presented in this paper with those from galaxy redshift distortions and the COBE normalization of a cold dark matter fluctuation spectrum for the flat model.

Figure 13. The critical overdensity for collapse as a function of the present value of the density parameter and redshift for the open and flat models.

Figure 14. The linear theory growth factor from a redshift z to the present as a function of the present value of the density parameter for the open and flat models.

Figure 15. The factors dependent on cosmology in the mass-temperature relation as a function of the present value of the density parameter and redshift for the open and flat models.

TABLE 1

LOW REDSHIFT SAMPLE

| Name | Z | F(2,10)[a] | EXOSAT kT(keV) | Einstein kT(keV) | GINGA kT(keV) | ASCA kT(keV) | AVE kT(keV) | Open ($\Omega_o$ = 0.5) $V_{sea}$[b] | n(>kT)[c] | REF |
|---|---|---|---|---|---|---|---|---|---|---|
| A2142 | 0.0894 | 7.5 | 11.0±1.9 | 9.4±1.0 | 8.68±0.20 | 8.8±0.6 | 8.74±0.19 | 1.63E8 | 0.00E+0 | 15,6,5,5,11 |
| A2029 | 0.0766 | 7.5 | | 7.8±1.2 | | 8.7±0.3 | 8.65±0.29 | 1.06E8 | 6.13E-9 | 19,5,11 |
| A401 | 0.0736 | 5.9 | | 7.8±1.0 | | 8.3±0.5 | 8.20±0.45 | 6.77E7 | 1.56E-8 | 3,5,11 |
| A1656 | 0.0232 | 32.0 | 8.0±0.3 | 8.3±0.9 | 8.21±0.09 | | 8.19±0.09 | 2.81E7 | 3.03E-8 | 3,6,5,10 |
| A754 | 0.0543 | 8.5 | 8.7±1.7 | 9.1±1.1 | 7.57±0.31 | 9.0±0.5 | 8.04±0.25 | 4.80E7 | 6.59E+8 | 3,6,5,2,11 |
| A2256 | 0.0581 | 5.2 | | 7.3±0.7 | 7.51±0.19 | 7.5±0.4 | 7.50±0.17 | 2.89E7 | 8.68E-8 | 3,5,2,11 |
| A399 | 0.0723 | 3.4 | | 5.8±1.3 | | 7.4±0.7 | 7.04±0.62 | 2.94E7 | 1.21E-7 | 3,5,11 |
| A3571 | 0.0397 | 11.5 | 7.6±1.1 | 10.9±2.9 | | 6.9±0.3 | 6.99±0.29 | 3.02E7 | 1.55E-7 | 13,6,9,11 |
| A478 | 0.0882 | 6.6 | 6.8±1.1 | 7.8±1.0 | 6.70±0.19 | 7.1±0.4 | 6.80±0.17 | 1.32E8 | 1.88E-7 | 19,6,5,18,11 |
| A3667 | 0.0556 | 6.7 | 5.3±1.6 | 6.5±1.6 | | 7.0±0.6 | 6.76±0.53 | 3.66E7 | 1.96E-7 | 3,14,5,11 |
| A3266 | 0.0590 | 5.9 | | 6.2±0.6 | | 7.7±0.8 | 6.74±0.48 | 3.61E7 | 2.23E-7 | 3,5,11 |
| A1651 | 0.0846 | 3.7 | | | | 6.3±0.5 | 6.30±0.50 | 5.20E7 | 2.51E-7 | 19,11 |
| A85 | 0.0556 | 6.4 | | 6.2±0.5 | | 6.1±0.2 | 6.11±0.19 | 3.42E7 | 2.70E-7 | 3,5,11 |
| A119 | 0.0442 | 3.0 | 5.1±0.9 | 5.9±1.0 | | 5.8±0.6 | 5.65±0.45 | 5.90E6 | 3.00E-7 | 3,6,5,11 |
| A3558 | 0.0479 | 4.2 | 3.8±2.0 | | 5.70±0.20 | 5.5±0.3 | 5.63±0.17 | 1.22E7 | 4.69E-7 | 3,5,4,11 |
| A1795 | 0.0630 | 5.3 | 5.1±0.5 | 5.8±0.5 | 5.34±0.11 | 6.0±0.3 | 5.42±0.10 | 3.73E7 | 5.51E-7 | 3,6,5,5,11 |
| A2199 | 0.0299 | 7.1 | 4.7±0.4 | 4.5±0.3 | 4.54±0.10 | | 4.54±0.09 | 6.63E6 | 5.78E-7 | 3,6,5,1 |
| A2147 | 0.0353 | 3.3 | 4.4±1.7 | 4.4±0.4 | | | 4.40±0.39 | 3.50E6 | 7.29E-7 | 15,6,5 |
| A3562 | 0.0502 | 3.5 | 3.8±0.9 | 4.2±1.2 | | | 3.94±0.72 | 1.07E7 | 1.01E-6 | 13,6,9 |
| A496 | 0.0328 | 5.7 | 4.7±0.9 | 3.9±0.2 | 3.91±0.06 | | 3.91±0.06 | 6.30E6 | 1.11E-6 | 3,6,5,18 |
| A3526 | 0.0103 | 11.2 | 3.6±0.4 | 3.9±0.2 | 3.54±0.13 | | 3.64±0.11 | 5.63E5 | 1.27E-6 | 3,6,5,18 |
| A1367 | 0.0213 | 3.5 | 3.5±0.5 | 3.7±0.3 | 3.50±0.18 | | 3.56±0.14 | 8.64E5 | 3.04E-6 | 3,6,5,1 |
| A1060 | 0.0124 | 4.4 | 3.3±0.4 | 3.9±0.4 | 3.27±0.05 | 3.1±0.2 | 3.27±0.05 | 2.43E5 | 4.20E-6 | 3,6,5,8,17 |
| C0336 | 0.0349 | 4.7 | 3.1±0.3 | 3.0±0.1 | 3.45±0.10 | | 3.22±0.07 | 5.70E6 | 8.32E-6 | 12,6,9,16 |
| Virgo | 0.0039 | 30.0 | 2.4±0.3 | 2.6±0.1 | 2.20±0.02 | | 2.21±0.02 | 1.35E5 | 8.49E-6 | 7,6,9,2 |

[a] $10^{-11}$ erg cm$^{-2}$ s$^{-1}$ in the 2 to 10 keV band
[b] $h^{-3}$ Mpc$^{3}$
[c] $h^{3}$ Mpc$^{-3}$
All fluxes are from Edge, Stewart, & Fabian (1992), all temperature errors are at the 90% confidence level.

References: 1. Arnaud (1994); 2 Arnaud & Evrard (1998); 3 den Hartog & Katgert (1996); 4 Day et al (1991); 5 David et al (1993); 6 Edge & Stewart (1991); 7 Federspiel, Tammann, & Sandage (1998); 8 Hatsukade (1989); 9 HA91; 10 Hughes et al (1993); 11 Markevitch et al (1998); 12 McNamara, Bregman, & O'Connell (1990); 13 Melnick & Moles (1987); 14 Piro & Fusco-Femiano (1988); 15 Struble & Rood (1987); 16 Tsuru (1993); 17 Tamura et al (1996); 18 Yamashita (1992); 19 Zabludoff, Huchra, & Geller (1990)

TABLE 2

HIGH REDSHIFT SAMPLE

| | | $F_{det}(0.3,3.5)$ ($10^{-13}$ erg cm$^{-2}$ s$^{-1}$) | | $L_{44}(2,10)$[a] $h_{50}^{-2}$ erg s$^{-1}$ | kT | Open ($\Omega_0$ = 0.5) | | Flat ($\Omega_0$ = 0.5) | |
|---|---|---|---|---|---|---|---|---|---|
| MS | Z | EMSS | ASCA | ASCA | (keV) | $V_{sea}$[b] | n(>kT)[c] | $V_{sea}$[b] | n(>kT)[c] |
| 0451.6 | 0.5392 | 9.5±1.7 | 12.4±0.2 | 30.6±0.6 | $10.3^{+0.9}_{-0.8}$ | 1.41E8 | 0.00E+0 | 1.80E8 | 0.00E+0 |
| 0015.9 | 0.5466 | 7.1±0.8 | 14.1±0.2 | 34.1±0.6 | 8.9±0.6 | 1.42E8 | 7.09E-9 | 1.81E8 | 5.57E-9 |
| 1008.1 | 0.3062 | 5.9±0.8 | 9.2±0.3 | 8.7±0.3 | $8.2^{+1.2}_{-1.1}$ | 6.86E7 | 1.41E-8 | 8.29E7 | 1.11E-8 |
| 1358.4 | 0.3290 | 12.2±2.1 | 11.0±0.3 | 10.2±0.3 | 6.9±0.5 | 9.13E7 | 2.87E-8 | 1.14E8 | 2.32E-8 |
| 1621.5 | 0.4274 | 3.4±0.6 | 5.2±0.2 | 7.5±0.3 | $6.6^{+0.9}_{-0.8}$ | 7.11E7 | 3.96E-8 | 8.96E7 | 3.20E-8 |
| 0353.6 | 0.320 | 6.3±1.3 | 7.8±0.3 | 6.8±0.3 | $6.5^{+1.0}_{-0.8}$ | 5.83E7 | 5.37E-8 | 7.09E7 | 4.31E-8 |
| 1426.4 | 0.320 | 4.4±0.6 | 5.3±0.2 | 4.6±0.2 | $6.4^{+1.0}_{-1.2}$ | 2.69E7 | 7.09E-8 | 3.24E7 | 5.72E-8 |
| 1241.5 | 0.549 | 4.2±1.1 | 4.7±0.3 | 10.7±0.6 | $6.1^{+1.4}_{-1.1}$ | 1.02E8 | 1.08E-7 | 1.30E8 | 8.81E-8 |
| 1147.3 | 0.303 | 3.0±0.7 | 5.3±0.2 | 4.1±0.2 | $6.0^{+1.0}_{-0.7}$ | 2.13E7 | 1.18E-7 | 2.55E7 | 9.58E-8 |
| 0811.5 | 0.312 | 2.6±0.5 | 3.5±0.2 | 2.6±0.2 | $4.9^{+1.0}_{-0.6}$ | 6.98E6 | 1.65E-7 | 8.34E6 | 1.35E-7 |
| 2137.3 | 0.313 | 19.3±2.6 | 20.9±0.4 | 15.3±0.3 | 4.9±0.3 | 1.27E8 | 3.08E-7 | 1.60E8 | 2.55E-7 |
| 0302.7 | 0.4246 | 3.8±0.4 | 4.2±0.3 | 4.8±0.3 | $4.4^{+0.8}_{-0.6}$ | 4.89E7 | 3.16E-7 | 6.10E7 | 2.61E-7 |
| 1224.7 | 0.3255 | 5.3±1.1 | 4.2±0.2 | 3.0±0.2 | $4.1^{+0.7}_{-0.5}$ | 1.64E7 | 3.36E-7 | 1.99E7 | 2.78E-7 |
| 1512.4 | 0.3726 | 4.5±1.0 | 4.5±0.3 | 3.8±0.3 | 3.4±0.4 | 3.47E7 | 3.97E-7 | 4.26E7 | 3.28E-7 |
| 2053.7 | 0.583 | 2.48±0.58 | 2.4±0.2 | 6.6±0.6 | $8.1^{+3.7}_{-2.2}$ | | | | |

[a] $\Omega_0 = 1$ [b] $h^{-3}$ Mpc$^3$ [c] $h^3$ Mpc$^{-3}$

All errors are at the 68% confidence level. Redshifts with four figures are from Carlberg et al (1996), those with three figures are from Gioia & Luppino (1994) except MS1241.5 is from Gioia (private communication).

TABLE 3

HIGH REDSHIFT SAMPLE SKY COVERAGE

| $F_{det}(0.3,3.5)$ ($10^{-13}$ erg cm$^{-2}$ s$^{-1}$) | Differential Solid Angle ($d\Omega_{surv}$) (ster) | Integral Solid Angle ($\Omega_{surv}$) (ster) |
|---|---|---|
| 2.50 | 0.0551 | 0.0551 |
| 2.78 | 0.0097 | 0.0648 |
| 3.34 | 0.0182 | 0.0830 |
| 4.01 | 0.0234 | 0.1064 |
| 4.81 | 0.0281 | 0.1345 |
| 5.77 | 0.0267 | 0.1613 |
| 6.93 | 0.0232 | 0.1845 |
| 8.31 | 0.0177 | 0.2022 |
| 10.0 | 0.0104 | 0.2126 |
| 11.9 | 0.0062 | 0.2188 |
| 14.3 | 0.00295 | 0.2218 |
| 17.3 | 0.00128 | 0.2230 |
| 20.7 | $5.5 \times 10^{-4}$ | 0.2236 |
| 24.8 | $1.5 \times 10^{-4}$ | 0.2237 |
| 29.7 | $3 \times 10^{-5}$ | 0.2238 |
| 35.7 | $3 \times 10^{-5}$ | 0.2238 |

TABLE 4

TEMPERATURE - BOLOMETRIC LUMINOSITY RELATION[a]

| Reference | <z> | A | $\alpha$ | kT @ $L_{44}$(bol) = 7[b] | Clusters |
|---|---|---|---|---|---|
| David et al (1993) | 0.066 | 2.94±0.14 | 0.297±0.004 | 5.24±0.20 | 104 |
| White, Jones & Forman (1997) | 0.057 | 2.76±0.08 | 0.33±0.01 | 5.25±0.05 | 86 |
| Henry & Sornig (1999) | 0.385 | 3.23±0.45 | 0.253±0.046 | $5.28^{+0.22}_{-0.31}$ | 29 |

[a]$kT = A\,[L_{44}(\mathrm{bol})]^{\alpha}$ for $H_0 = 50$ km $s^{-1}$ $Mpc^{-1}$, $\Omega_0 = 1$, and kT in keV
[b]errors allow for correlated errors on A and $\alpha$, ie –1 $\sigma$ on A is associated with +1 $\sigma$ on $\alpha$ and vice-versa.

TABLE 5

DETERMINATION OF THE MASS – TEMPERATURE PARAMETER $\beta_{TM}$

| Method | $\beta_{TM}$ | $\Omega_0$ | $\Omega_b$ | $\lambda_0$ | h | $\sigma_8$ | Objects | Ref |
|---|---|---|---|---|---|---|---|---|
| Hydrodynamic Simulation | 1.26±0.03 | 1.0 | 0.10 | 0 | 0.50 | 0.59 | 22 | 1 |
| Hydrodynamic Simulation | 1.17 | 1.0 | 0.10 | 0 | 0.50 | 0.59,0.63 | 42 | 2 |
| Hydrodynamic Simulation | 0.88 | 0.2 | 0.10 | 0, 0.8 | 0.50 | 1.00 | 16 | 2 |
| Hydrodynamic Simulation | 1.15±0.05 | 1.0 | 0.05 | 0 | 0.50 | 0.50 | 20 | 3 |
| Hydrodynamic Simulation | 1.07±0.04 | 0.3 | 0.04 | 0.7 | 0.70 | 1.05 | 30 | 4 |
| Hydrodynamic Simulation | 1.29 | 1.0 | 0.06 | 0 | 0.50 | 1.05 | 25 | 5 |
| Hydrodynamic Simulation | 1.36 | 1.0 | 0.10 | 0 | 0.65 | 0.67 | 25 | 5 |
| Hydrodynamic Simulation | 1.36 | 1.0 | 0.075 | 0 | 0.50 | 0.70 | 25 | 5 |
| Hydrodynamic Simulation | 1.31 | 0.4 | 0.06 | 0 | 0.65 | 0.75 | 25 | 5 |
| Truncated Isothermal Sphere | 1.16 | 1.0 | | 0 | | | | 6 |
| Gravitational Lens | 1.02±0.11 | 1.0 | | 0 | 0.50 | | 8 | 7 |
| Velocity Dispersion | 1.21±0.35 | | | | 1.0 | | 27 | 8 |
| Spatially Resolved Temperature | $0.82^{+0.31}_{-0.18}$ | | | | 1.0 | | 11 | 8 |

References: 1. Evrard (1990); 2. Evrard, Metzler, & Navarro (1996); 3. Pen (1998); 4. Eke, Navarro, & Frenk (1998); 5. Bryan & Norman (1998); 6. Shapiro, Iliev, & Raga (1998); 7. Hjorth, Oukbir, and van Kampen (1998); 8. Horner, Mushotzky, & Scharf (1999)

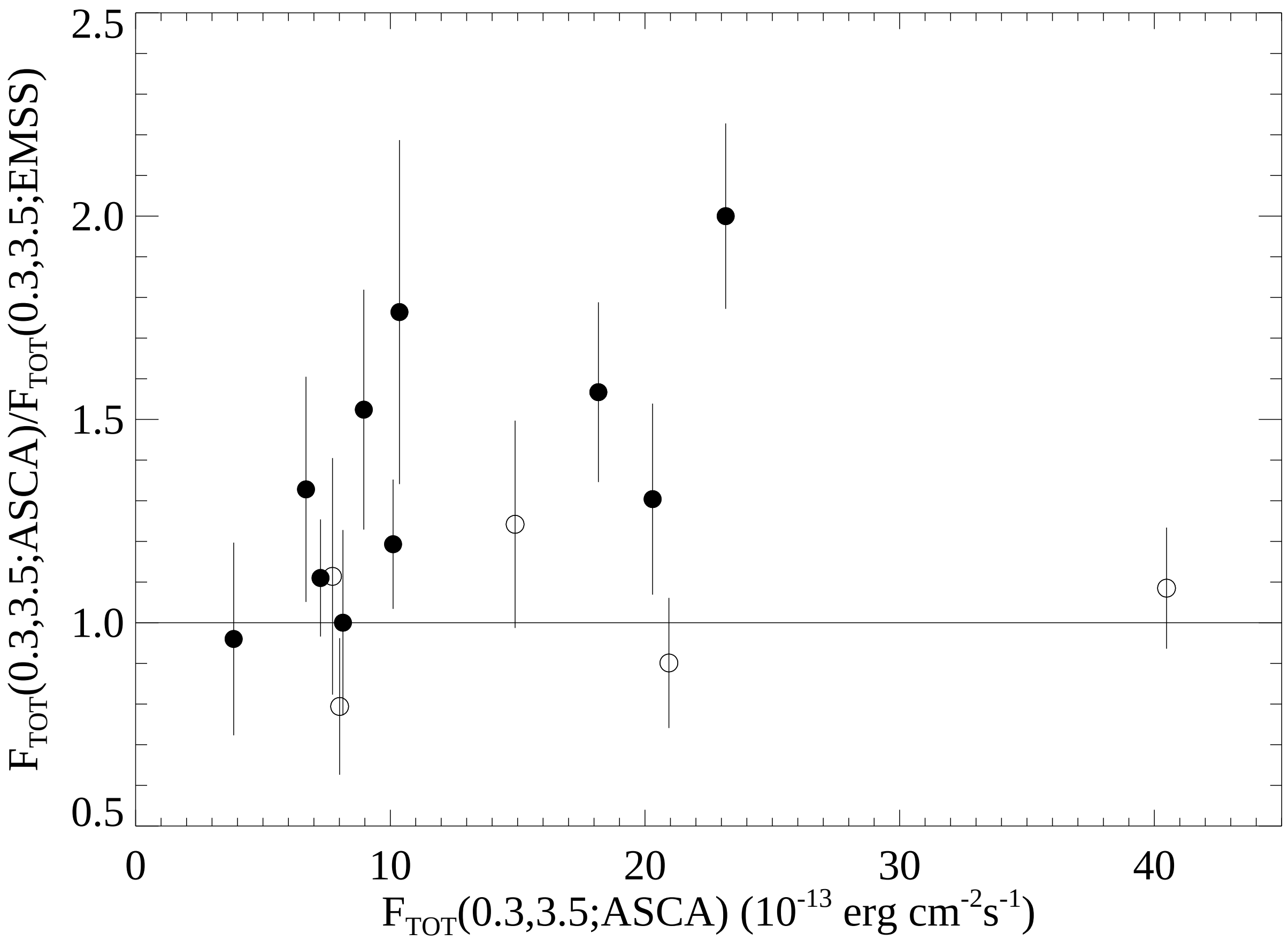

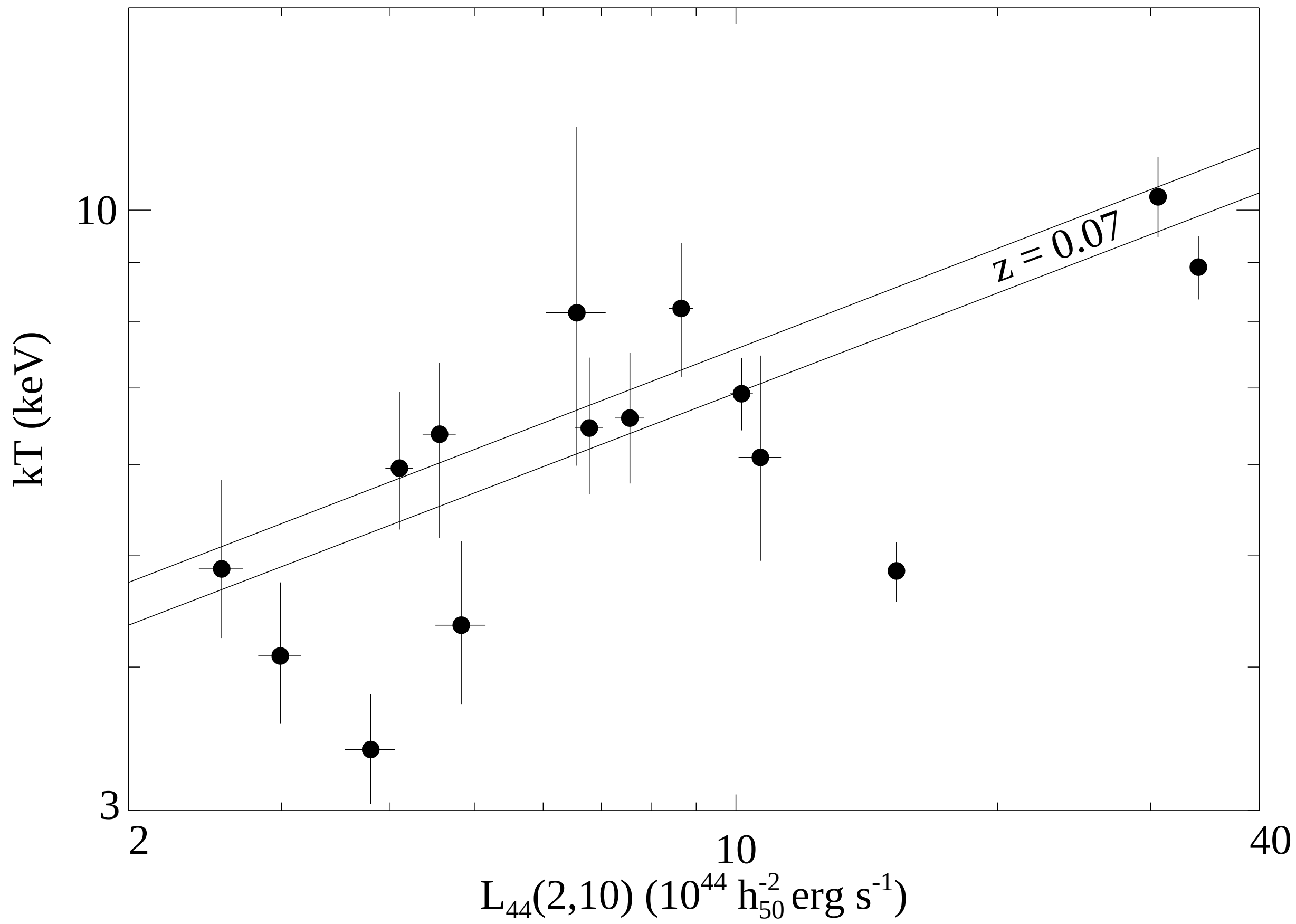

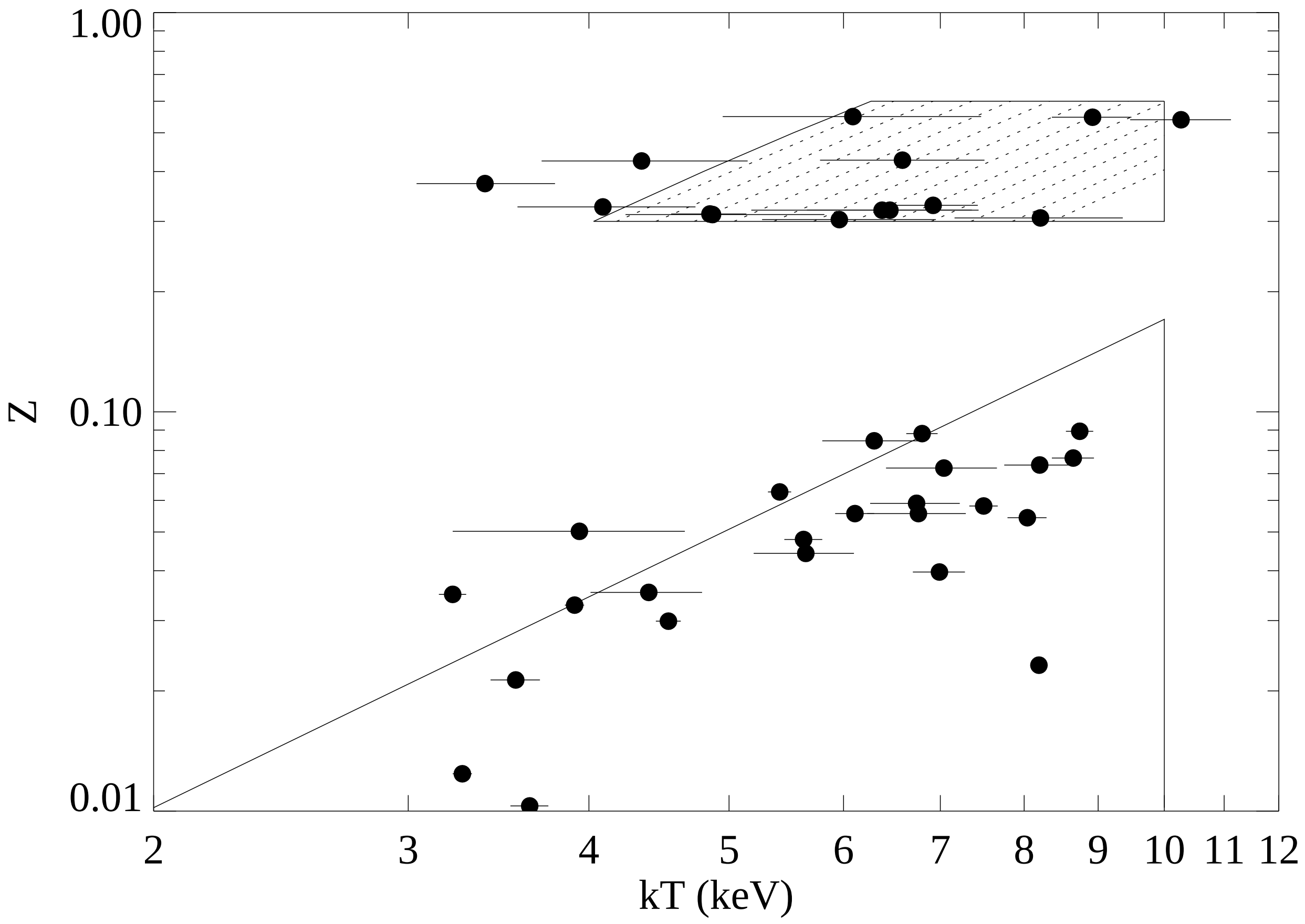

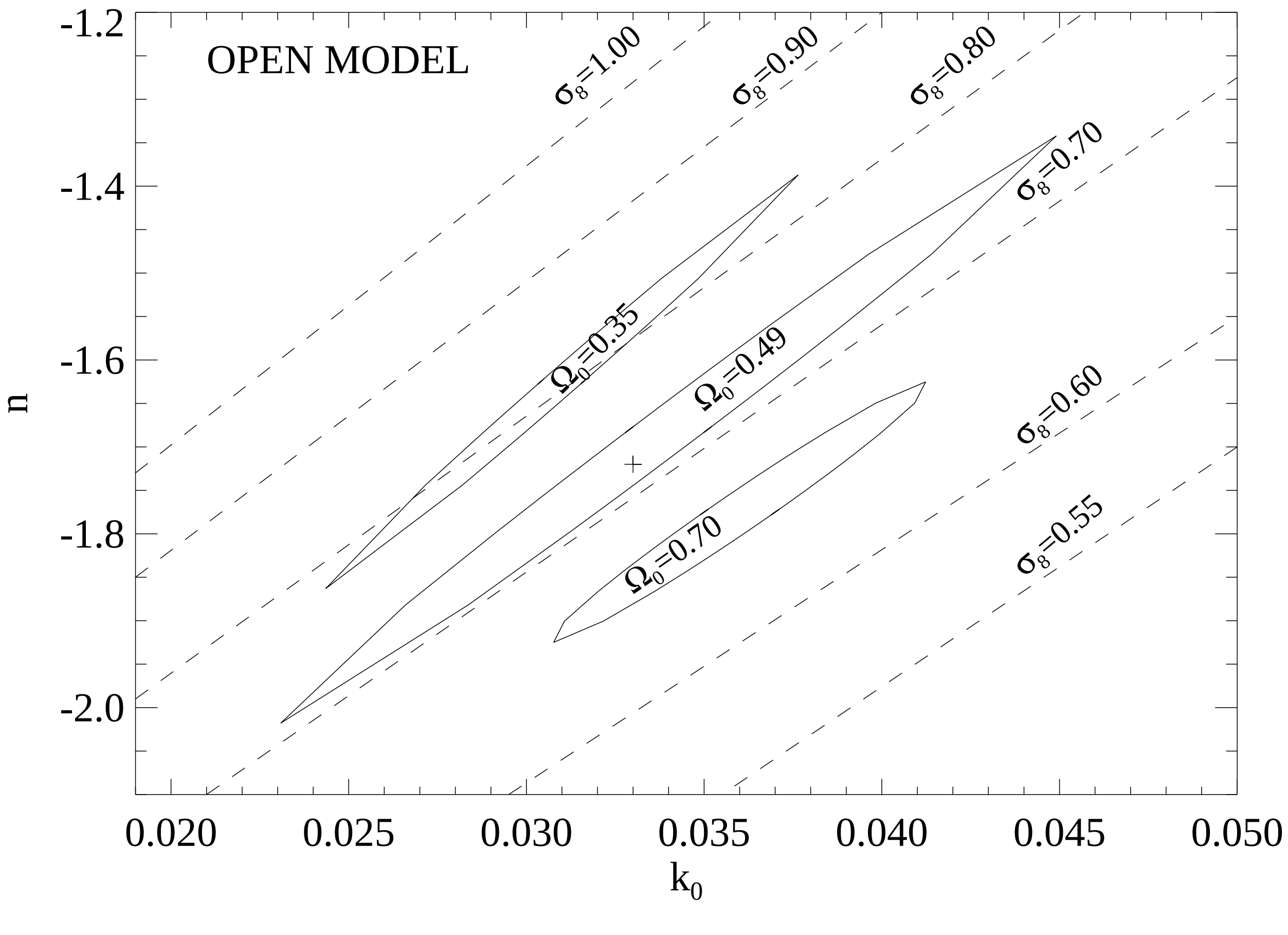

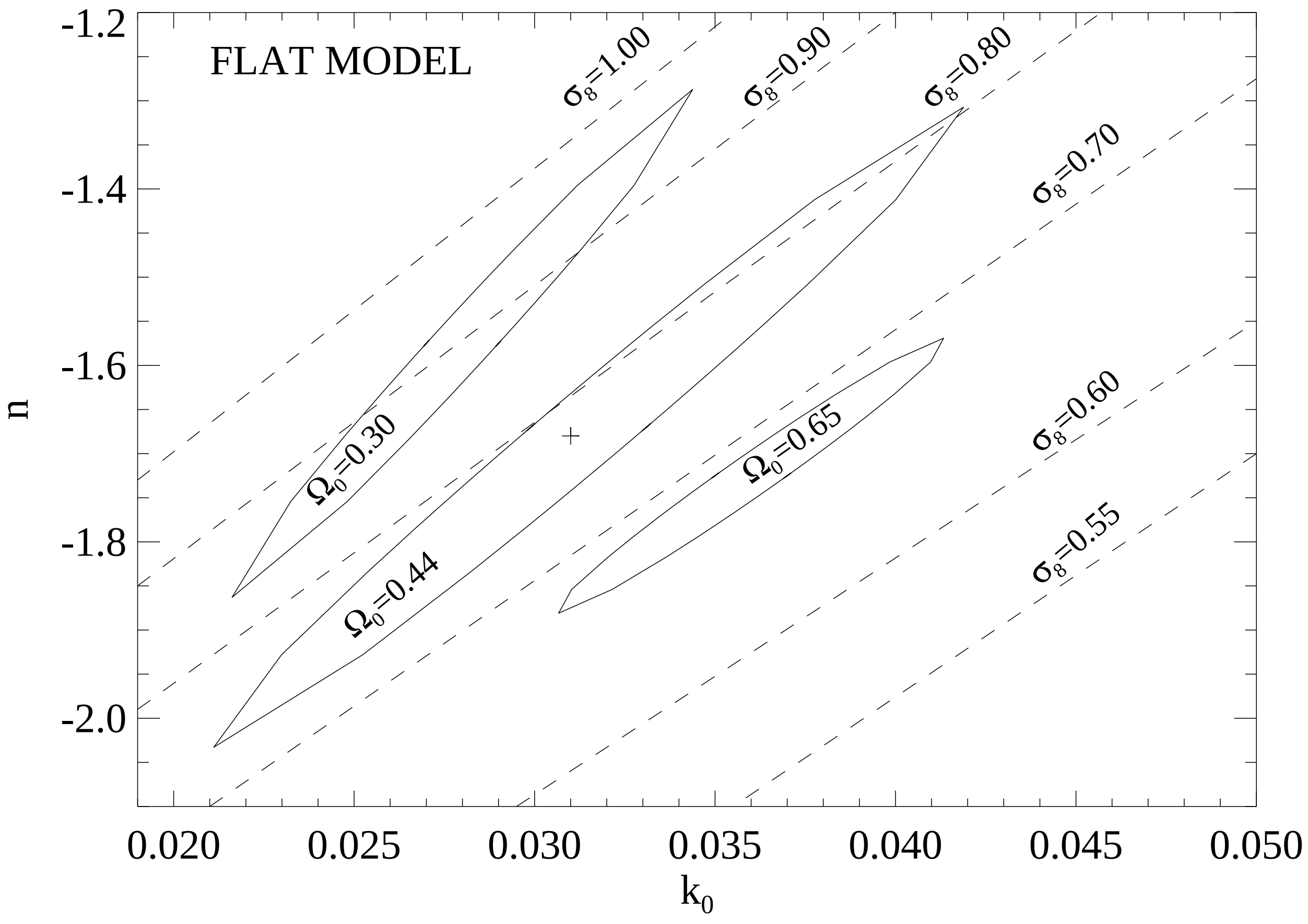

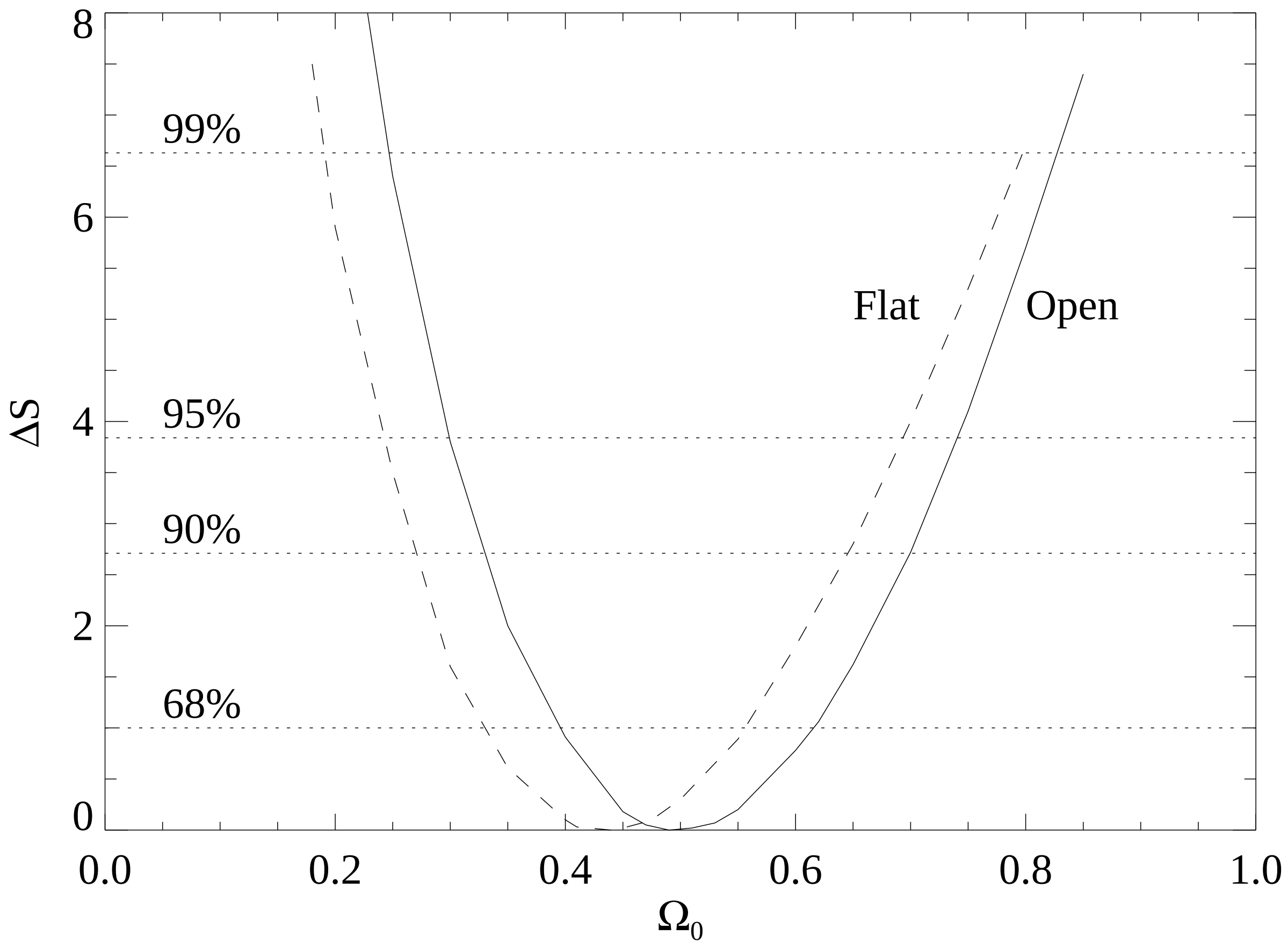

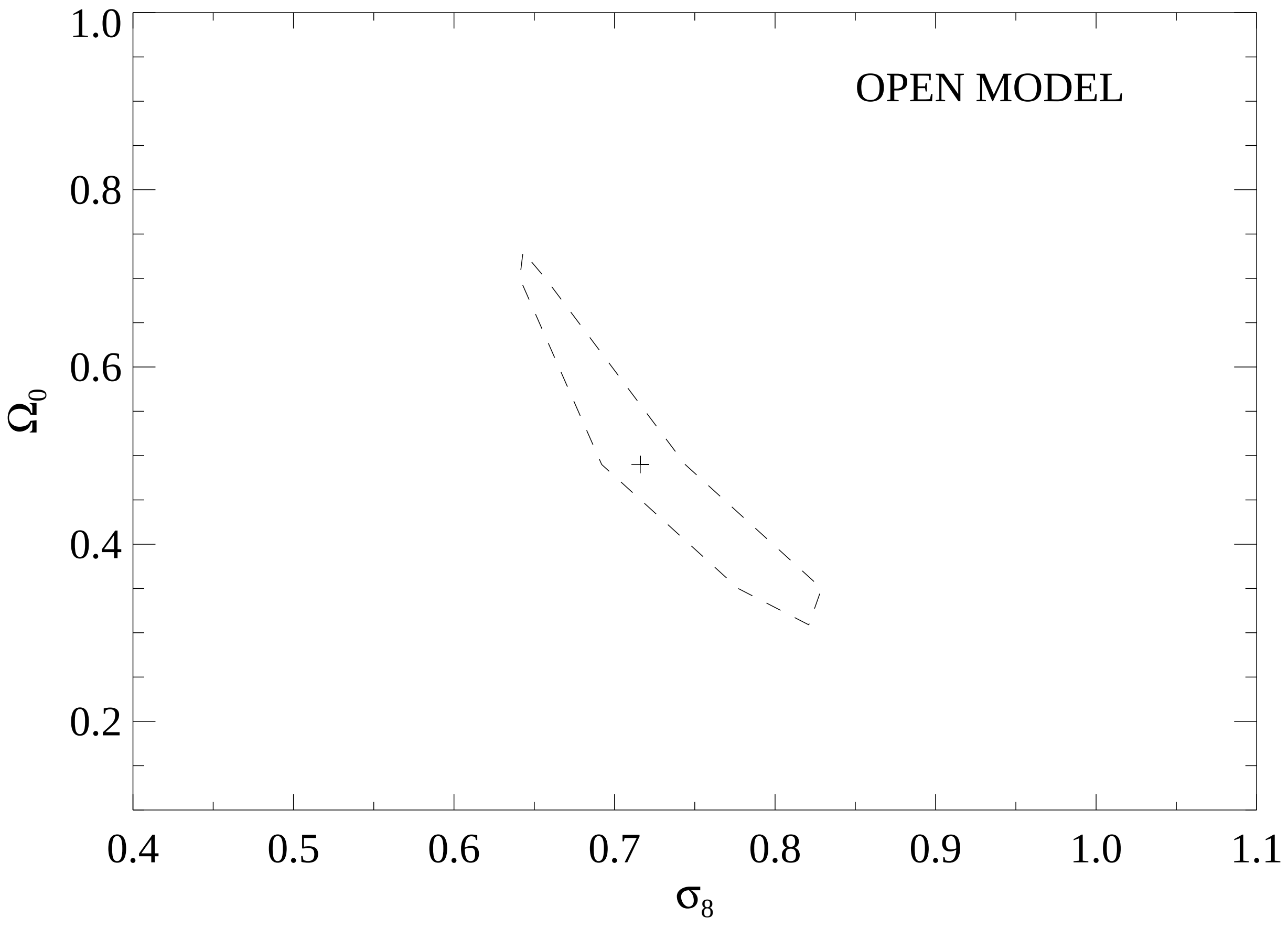

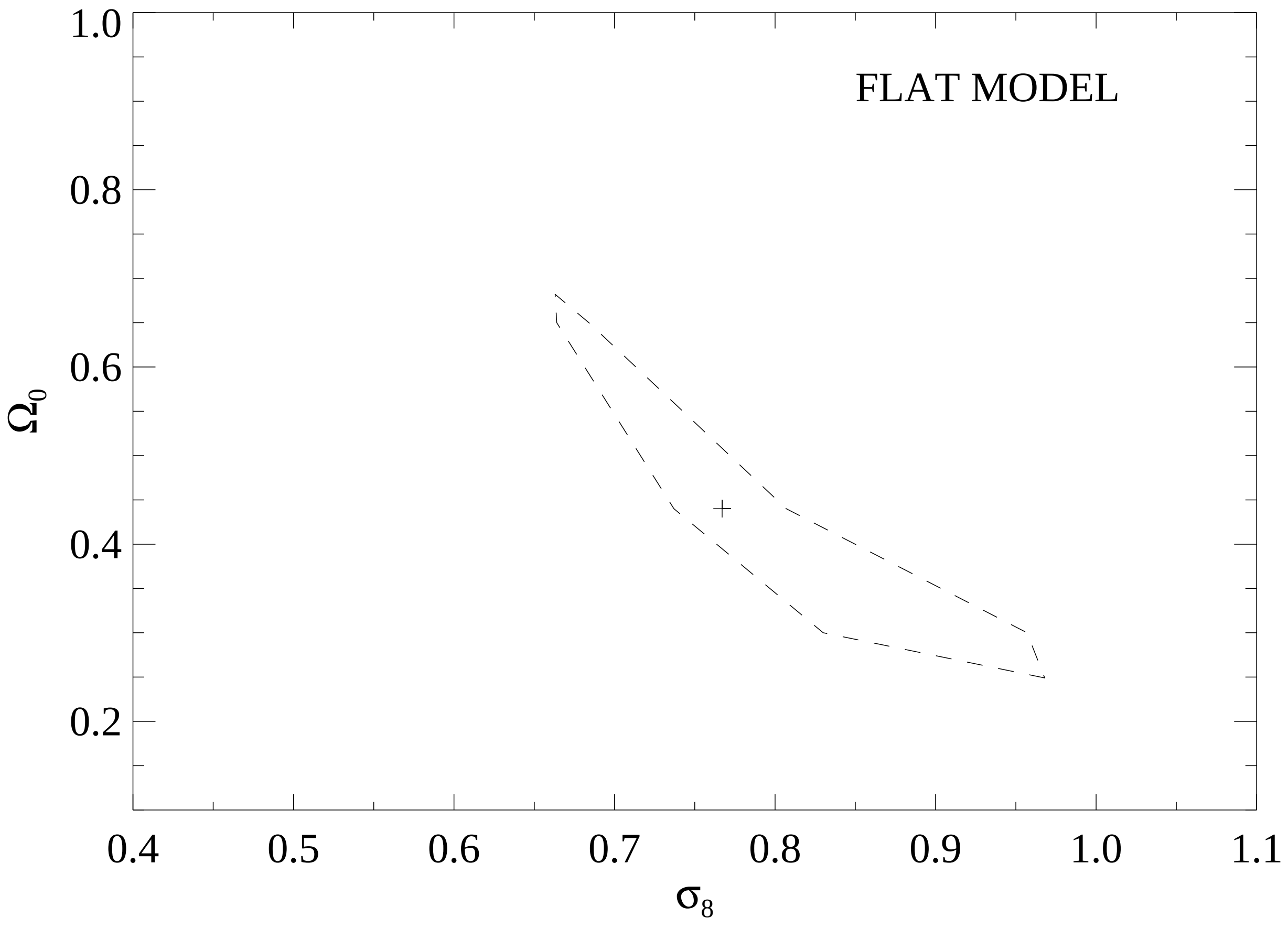

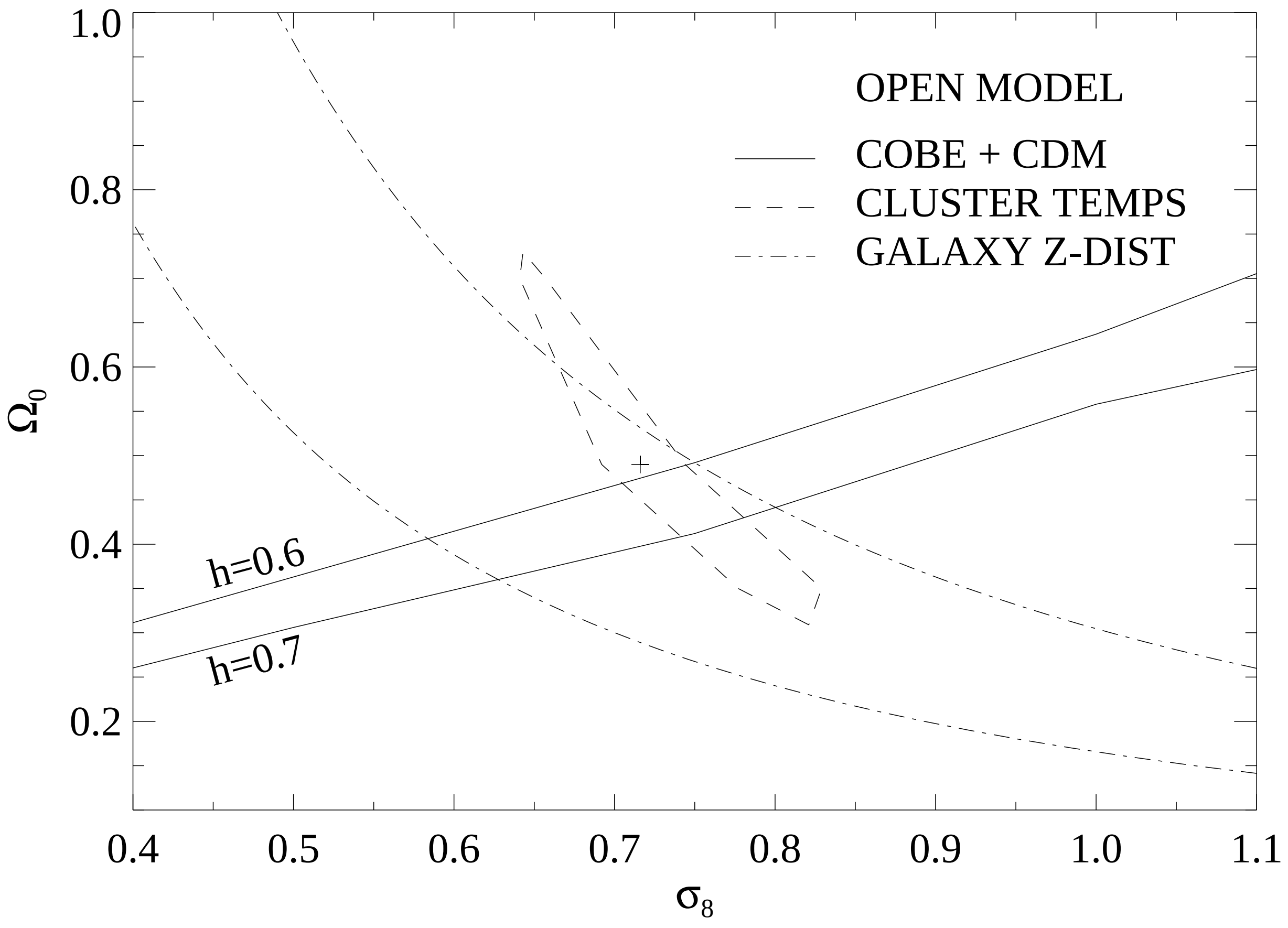

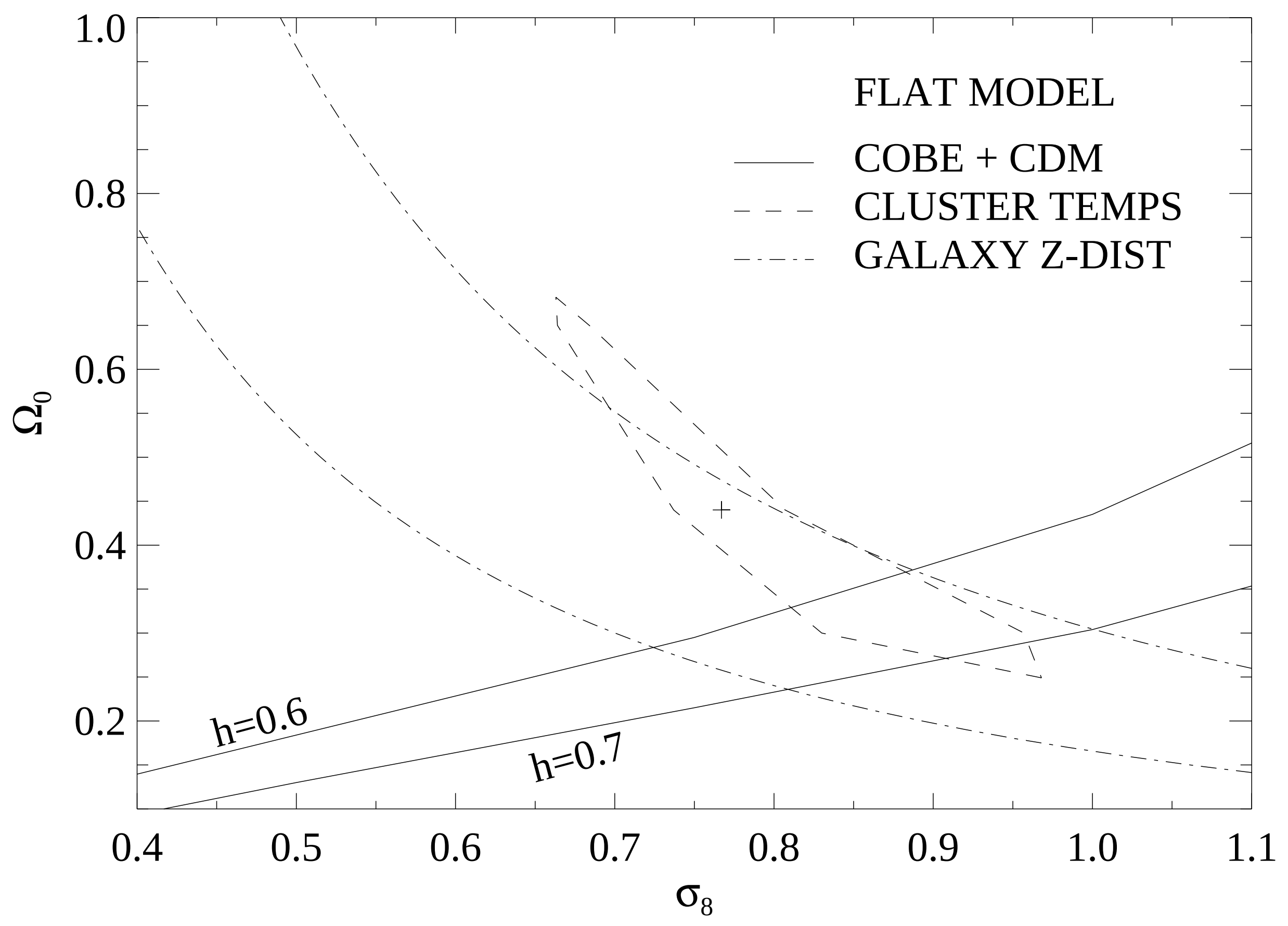

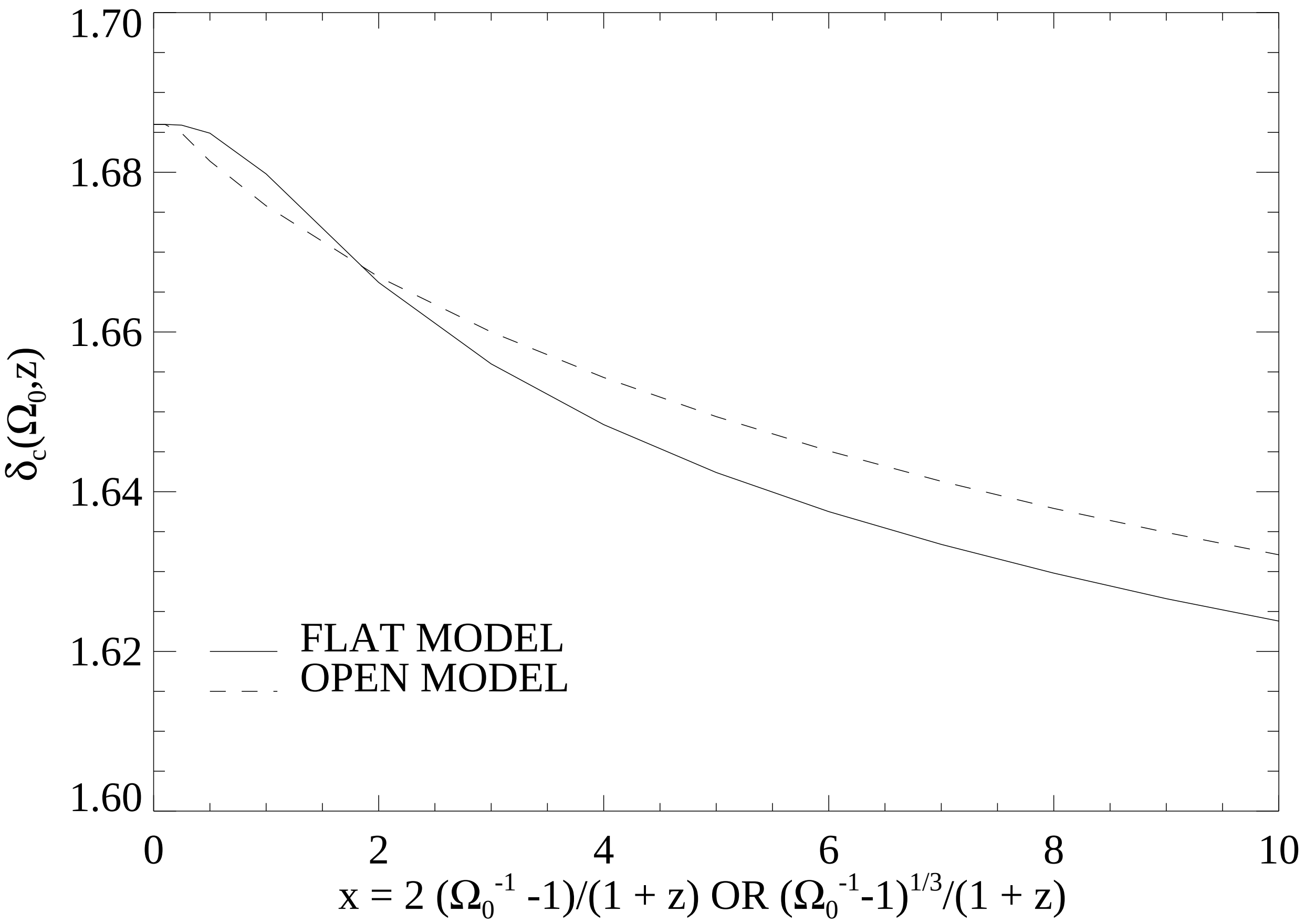

1.70
1.68
1.66
1.64
1.62
1.60
$\delta_c(\Omega_0,z)$
0
2
4
6
8
10
x = 2 ($\Omega_0^{-1}$ -1)/(1 + z) OR ($\Omega_0^{-1}$-1)$^{1/3}$/(1 + z)
FLAT MODEL
OPEN MODEL

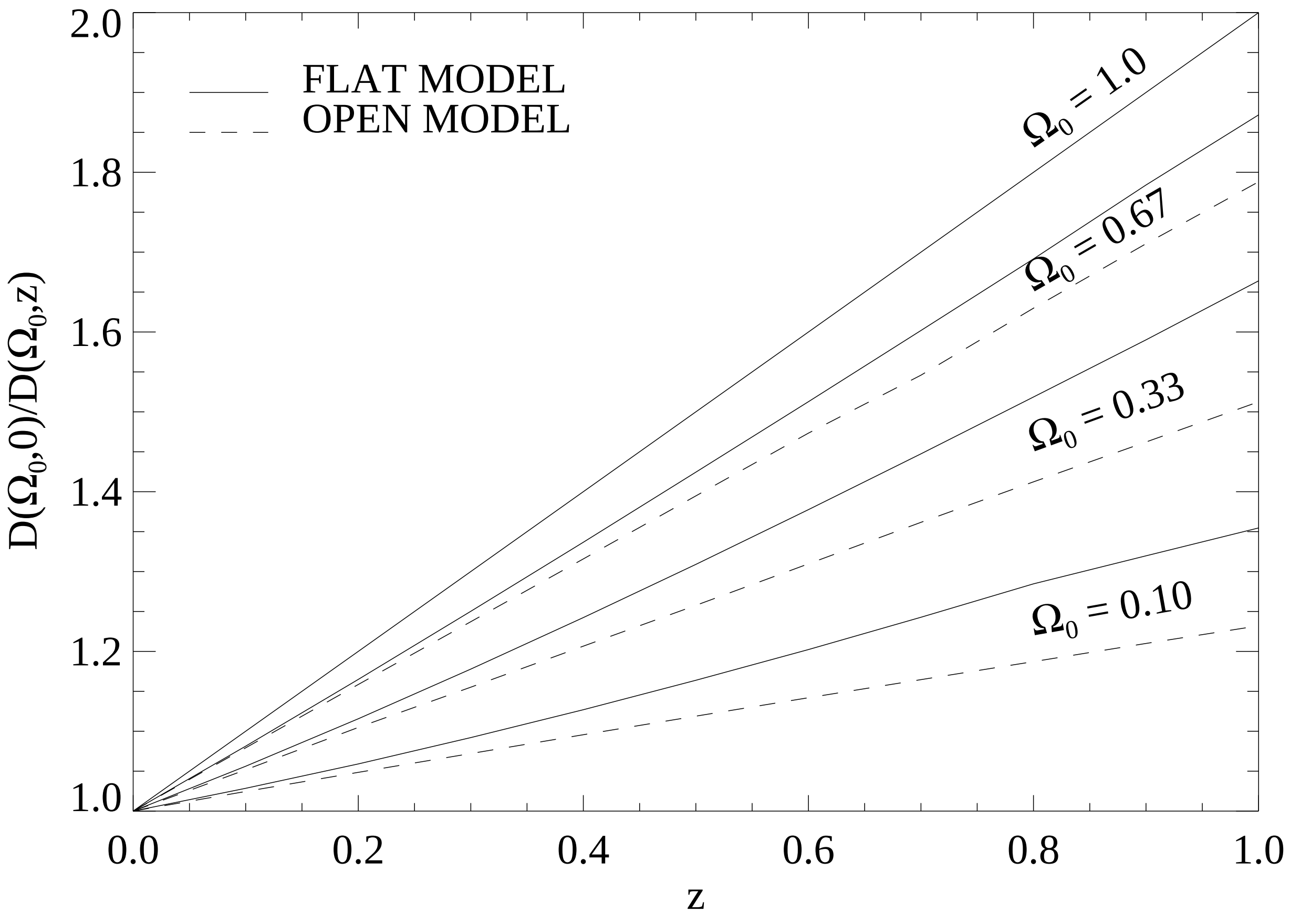

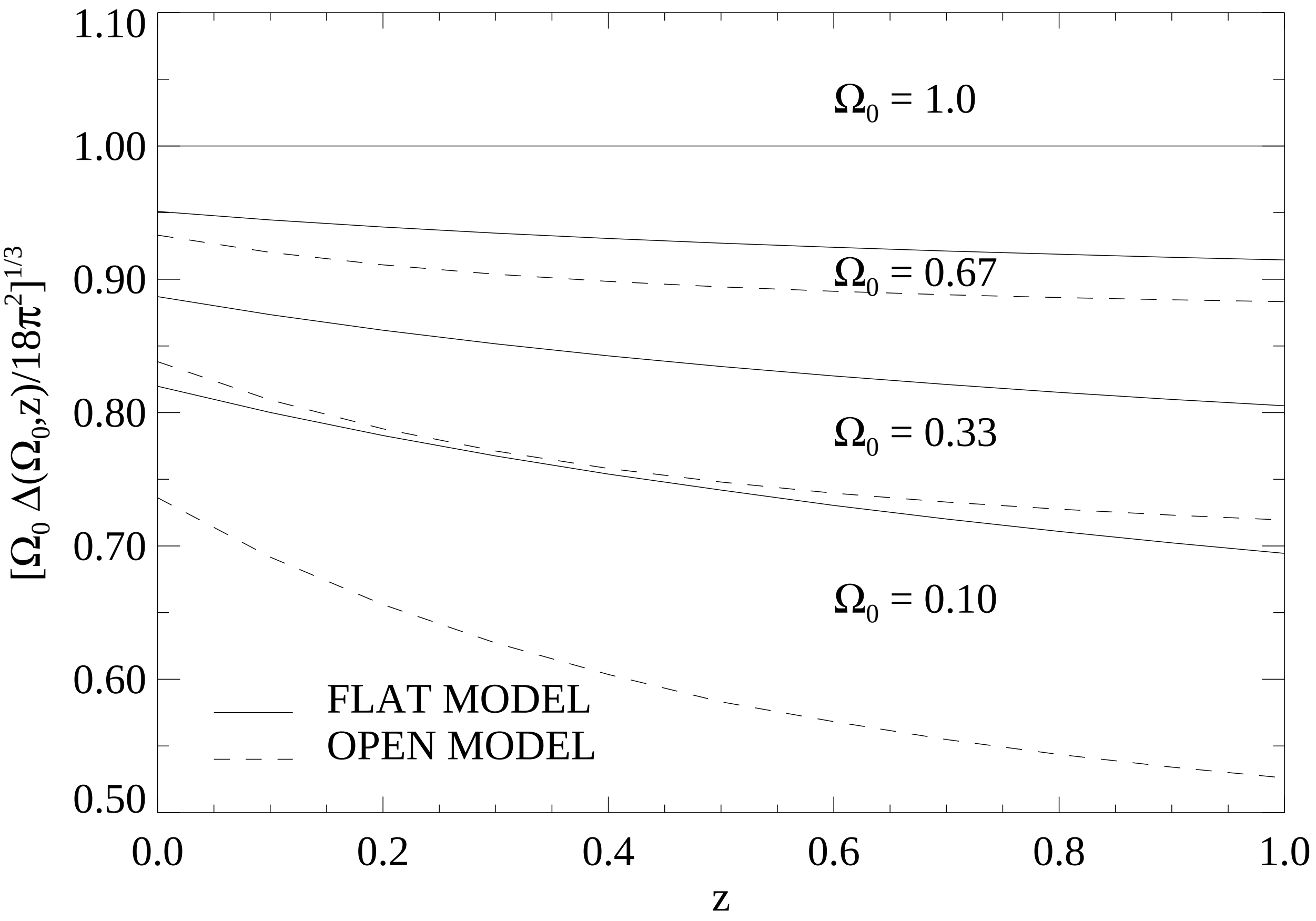